\documentclass[a4paper,fleqn]{cas-sc}

\usepackage[numbers]{natbib}
\usepackage{graphicx}
\usepackage{amsmath}
\usepackage{algorithm}
\usepackage{algpseudocode}

\usepackage[version=4]{mhchem}
\usepackage{siunitx}
\usepackage{longtable,tabularx}
\usepackage{subcaption}
\usepackage{tikz}
\usepackage{pgfplots}
\usetikzlibrary{patterns}
\def\tsc#1{\csdef{#1}{\textsc{\lowercase{#1}}\xspace}}
\tsc{WGM}
\tsc{QE}

\begin{document}
\let\WriteBookmarks\relax
\def\floatpagepagefraction{1}
\def\textpagefraction{.001}

\shorttitle{An efficient GPU-based $h$-adaptation framework via linear trees for the flux reconstruction method}    

\shortauthors{Lai Wang}  

\title [mode = title]{An efficient GPU-based $h$-adaptation framework via linear trees for the flux reconstruction method}  

\tnotemark[1] 


%

\author[1]{Lai Wang}

\cormark[1]


\ead{bx58858@umbc.edu}

\credit{Conceptualization, Methodology, Software, Writing} 

\affiliation[1]{ 
            city={Gaithersburg},
            postcode={20878}, 
            state={MD},
            country={USA}}
\author[2]{Freddie Witherden} 


\ead{fdw@tamu.edu}
 
\credit{Conceptualization}

\affiliation[2]{organization={Department of Ocean Engineering},
            addressline={Texas A\&M University}, 
            city={College Station},
            postcode={77843}, 
            state={TX},
            country={USA}}
\author[2,3]{Antony Jameson} 


\ead{antony.jameson@tamu.edu}
 
\credit{Conceptualization}

\affiliation[3]{organization={Department of Aerospace Engineering},
            addressline={Texas A\&M University}, 
            city={College Station},
            postcode={77843}, 
            state={TX},
            country={USA}}

\cortext[1]{Corresponding author} 

\begin{abstract}
In this paper, we develop the first entirely graphic processing unit (GPU) based $h$-adaptive flux reconstruction (FR) method with linear trees. 
The adaptive solver fully operates on the GPU hardware, using a linear quadtree for two dimensional (2D) problems and a linear octree for three dimensional (3D) problems.
We articulate how to efficiently perform tree construction, 2:1 balancing, connectivity query, and how to perform adaptation for the flux reconstruction method on the GPU hardware. 
As a proof of concept, we apply the adaptive flux reconstruction method to solve the inviscid isentropic vortex propagation problem on 2D and 3D meshes to demonstrate the efficiency of the developed adaptive FR method on a single GPU card. Depending on the computational domain size, acceleration of one or two orders of magnitude can be achieved compared to uniform meshing. 
The total computational cost of adaption, including tree manipulations, connectivity query and data transfer, compared to that of the numerical solver, is insignificant. 
It can be less than  2\% of the total wall clock time for 3D problems even if we perform adaptation as frequent as every 10 time steps with an explicit 3-stage Runge--Kutta time integrator.
\end{abstract}




\maketitle

\section{Introduction} 
The computational fluid dynamics (CFD) 2030 vision study stated that ``An engineer/scientist must be able to generate, analyze, and interpret a large ensemble of related simulations in a time-critical period (e.g., 24 hours), without individually managing each simulation, to a pre-specified level of accuracy and confidence"~\cite{slotnick2014cfd}. To this end, we would like to reduce the human intervention as much as possible in all stages of CFD simulations, including mesh generation, simulation, and data analyzing. 

A Cartesian-grid-based solver can be a powerful tool for  automated CFD simulations since it can almost eliminate human intervention in meshing and it has more flexibility to deal with moving and deforming objects. There are two major challenges of utilizing a Cartesian grid for wall-bounded CFD simulations, namely, (a) how to resolve wall boundaries accurately with/without turbulence modeling and (b) how to perform automated adaptive mesh refinement (AMR), including initial meshing and mesh adaptation for CFD solvers. 
To address the first issue,  the immersed boundary method~\cite{peskin2002immersed} has been arguably the most popular method for low-to-medium Reynolds numbers. The cut-cell approach~\cite{wang1998quadtree, ingram2003developments} and the extended finite element method~\cite{moes1999finite, kummer2021bosss} can provide more accurate resolution of the wall boundaries with additional cost to perform the boundary intersections and severe stiffness increase could be introduced by small corners. A Cartesian grid can also be helpful to serve as the background grid in the overset grid method~\cite{steger1983chimera, crabill2018parallel} to handle high Reynolds number flows with near-wall body-fitted meshes and wall models~\cite{galbraith2023hlpw}. However, more efforts will then be needed to automate the mesh generation for the boundary layer, such as extruding the surface mesh in wall normal direction, and it will be more challenging when the body starts to deform.
Without AMR one would have to manually refine the mesh resolution in the near-body region and weak region. A nonconforming Cartesian mesh is normally required to reduce the element count due to topological  requirements. Moreover, if one is to study the maneuver of any vehicle, the enlarged refined region to account for the motion would lead to an overwhelming total number of elements and hence a waste of computational resources. 

The last decade has witnessed the revolution of CFD embarked by the development of high-order methods, such as discontinuous Galerkin methods~\cite{cockburn2012discontinuous} and flux reconstruction methods~\cite{huynh2007flux}, which are better suited for parallel computing~\cite{witherden2014pyfr} and the development of modern accelerators, such as the graphic process unit (GPU). Discontinuous finite element type high-order methods are compact in nature and they have been found to be more efficient in scale-resolving simulations than their low-order counterparts~\cite{vermeire2017utility}. The compactness of these methods makes them great candidates to work with AMR methodologies for Cartesian grids in the sense that less number of elements will be needed to achieve the same resolution compared to other non-compact high-order methods such as the finite volume methods and the cost of adaptation will hence be lower. In addition,  the complexity of building up the high-order stencil on a non-uniform mesh for high-order finite difference methods will be overwhelming. GPU-based solvers have become more and more popular due to the tremendous speedup they offer compared to traditional CPU-based solvers.  In foreseeable future, utilizing the computational power of GPU cards would be more affordable compared to CPUs for massive problem sizes. There are developments done regarding GPU-based adaptive solvers using octrees. Many of them relied on the CPU hardware to perform the mesh adaptation and information would be sent to GPU from CPU~\cite{wahib2015data}. This streamline will magnify the bottleneck of data transfer between CPU and GPU. Hence, we would strive to develop efficient high-order CFD solvers with AMR that entirely operate on the GPU cards. We are interested in discontinuous finite element methods in particular due to the fact that elements will only need to communication through interface fluxes instead of relying on continuous values on element vertices as continuous Galerkin method does. 

Tree structures are widely used in AMR~\cite{BursteddeWilcoxGhattas11}. Specifically, one can use a quadtree for two dimensional problems and an octree for three dimensional problems. There are two prevalent approaches, namely, top-bottom and bottom-up trees.
The top-bottom approach is intuitively straightforward in the sense that one starts from the root node and gradually performs refinement until satisfaction and its implementation is commonly based on pointers. However, a plain pointer-based implementation do not have great data locality. A linear tree (quadtree/octree) is ubiquitous in that only a linear array is used to store the leaf nodes of the tree. And each leaf node is usually represented by a unique ID, such as the Morton ID, which encodes all the information, namely, coordinates and depth, of the represented leaf node. The bottom-up tree construction normally starts with spatial distribution of points that we are interested in, such as a human skull surface mesh we would like to render or a surface mesh of the airplane we would like to perform CFD studies. These points are immediately encoded as scattered octants. One needs to perform tree completion such that the resulting octants will continuously fill the domain with no holes and overlaps and all initial points are encapsulated in the finest octants or leaf nodes. In computational graphics, researchers have so far developed GPU-based octree data structure which can perform the bottom-up tree completion. However, when used in CFD, a tree needs to be 2:1 balanced. 2:1 balance refers to the fact that the depth difference of two adjacent octants cannot exceed 2 and it is needed for an easy enforcement of conservation. A meshless method could be an exception. One example is that in~\cite{jambunathan2017advanced}, Jambunathan and Levin developed a meshless Monte-Carlo method using an unbalanced octree on GPUs, where the initial tree was built on CPU first. In the work of Pavlukhin and Menshov~\cite{pavlukhin2019gpu}, an entirely GPU-based tree was implemented. And information of the 3-level of tree nodes were stored, two of which are interior nodes.  No enough information was provided on 2:1 balancing of the tree. The approach is overall different from the classic linear tree.
 The challenge of 2:1 balancing is to efficiently find the neighbors of the any given octant in the tree. 
Sundar et al.~\cite{sundar2008bottom} proposed the idea of using searching keys to to perform binary search in the sorted tree for efficient balancing and the local balancing on CPU hardware will rely on hash tables to reduce the duplicates of octants for optimal performance. On GPU hardware with massive parallelism, a thread-safe hash table with no locks is not readily available.  In this paper, we are going to articulate how one can performance efficient linear tree completion and tree balancing for GPU hardware without the use of any hash tables.

We argue that a tree of octants is not functionally equivalent to a mesh of elements for CFD simulations.
Finding connectivity among different entities in the mesh, such as elements, faces, points, is as important as the tree manipulation algorithms for finite element methods. On massively parallel hardware, the absence of efficient thread-safe hash tables is to complicate the conversion from a tree to a mesh that is usable for numerical methods. We will explain in this paper what makes discontinuous finite element methods great candidates to work with linear trees for efficient AMR on GPU hardware.

Mesh adaptation ($h$-adaptation) and polynomial adaptation ($p$-adaptation) methods are widely used for high-order discontinuous methods. Polynomial adaptation is commonly praised for its ease of implementation compared to mesh adaptation  and the great speedup that it can offer~\cite{wang2020dynamically}. However, things will be quite different for GPU computations. For massive parallelism with graphic cards, minimization of thread divergence is warranted for performance. Generally, there are two major types of calculation for any high-order discontinuous finite element methods, namely, element-related and interface-related calculations.  For mesh adaptation with linear trees, regardless of the depth of any octant in the tree, element-related calculations will be identical for all voxel elements (quadrilateral elements in 2D and hexahedral elements in 3D) including interpolations to the interfaces. Only conforming and nonconforming interior interfaces and boundary faces will need different procedures of common flux calculations and they are resolution-indifferent. For polynomial adaptation, the numerical discretization operators will be completely polynomial-degree dependent. Moreover, optimization of the data layout and shared memory utilization, could be very different and special treatment will be needed for high polynomial degree since the shared memory would not be big enough to fit the entire numerical operator, such as the differentiation matrix and projection matrices for data transfer in adaptation. The more polynomial degrees are used, the more significant the divergence would be. Therefore, mesh adaptation appears to be more attractive for GPU hardware for optimal performance. 

\textit{Contributions.} In this paper, we report a GPU-based $h$-adaptive flux reconstruction method with linear trees.  This is the first paper that elaborates the novel algorithms for tree completion and tree balancing and how to efficiently use linear tree for discontinuous finite element methods on GPU hardware. We provide enough algorithm/coding details for the readers.  The developed method has shown significant speedup for long distance vortex propagation and the speedup can be as many as 49 times. For 3D simulations, the adaptation cost including tree manipulations, face query and data transfer, is only around 2\% of the overall computational cost even though we perform adaptation as frequent as every 10 times with explicit time integrators. This work paves the way to further development of automated Cartesian-grid-based CFD solver which can utilizing massively parallel  hardware. 
 
\textit{Organizations.} The remainder of this paper is organized as follows. In Section~\ref{sec:background}, we introduce the flux reconstruction method and linear trees briefly. Section~\ref{sec:tree_manipulations} documents in detail how to construct and balance a linear tree entirely on GPU hardware. And we also illuminate how to build the face connectivity for the flux reconstruction method. In Section~\ref{sec:adaptive_fr}, we articulate how to perform adaptation and how data should be transferred in the adaptation procedure. We will present the numerical experiments in Section~\ref{sec:numerical_exp}. Section~\ref{sec:summary} summarizes this work.

\section{Background}\label{sec:background}

\subsection{The flux reconstruction method}
The flux reconstruction method was originally proposed by Huynh~\cite{huynh2007flux} for tensor product elements. 
We remark that for high-order collocation methods, the nested inner resolution has a great potential of making the right hand 
side evaluation dominate the total computational cost such that the cost of mesh adaptation can be significantly smaller 
compared to low order methods. This has previously been observed with a $p$-adaptive flux reconstruction method 
on unstructured grids~\cite{wang2020dynamically} where as the polynomial degree increases, the adaptation overhead gets smaller.

Considering the following 3D conservation law  (2D problems are treated similar)
\begin{equation}
	\frac{\partial u}{\partial t} + \frac{\partial f}{\partial x} + \frac{\partial g}{\partial y} + \frac{\partial h}{\partial z} = 0
\end{equation}
on a Cartesian grid, we can calculate the spatial derivatives dimension by dimension as illuminated by Huynh~\cite{huynh2007flux}. Taking $f$ as an example, we first construct the corrected  flux polynomial $\hat{f}$ as 
\begin{equation}\label{eq:}
	\hat{f} = f (\xi) + g_L (\xi) (\hat{f}(-1) - f(-1)) + g_R(\xi) (\hat{f}(1)-f(1)),
\end{equation}
in the local coordinate system, where the local flux polynomial $f(\xi)$ is represented using a Lagrangian polynomial as 
\begin{equation}
	f(\xi) = \sum_{i=1}^{k+1} l_i(\xi) f_i,
\end{equation}
and $k$ is the polynomial degree of $l_i(\xi)$.   $\hat{f}(\pm 1)$ are numerical fluxes at the interfaces,
 which are  referred to as the common fluxes. And a Riemann solver is normally used for their calculations. 
 $g_L$ and $g_R$ are left and right correction functions, which are symmetric with respect to $\xi = 0$.
The derivative of the corrected flux polynomial can be expressed as 
\begin{equation}
	\hat{f}'= \sum_{i=1}^{k+1} l'_i(\xi) f_i + g'_L(\xi)(\hat{f}(-1)-f(-1)) + g'_R(\xi) (\hat{f}(1)-f(1)).
\end{equation}

In terms of implementation, these differentiation operators are normally pre-computed. In our GPU implementation, we store the local differentiation operator of $f$ as well as $g'_L(\xi)$ in the constant memory. Radau polynomials are used in this work~\cite{huynh2007flux} to recover the discontinuous Galerkin method.  A finite difference approach was introduced by Wang and Yu~\cite{wang2018compact} to offer a different perspective that one can organize the local differentiation and correction from the common flux into one operator.
The derivative with respect to $x$ can be computed as 
\begin{equation}
	\frac{\partial f}{\partial x} = \frac{\partial f}{\partial \xi} \frac{\partial \xi}{\partial x}.
\end{equation}
Even though the mesh in AMR is non-uniform,  $\Delta x$ of every element is not to be explicitly stored since it can be inferred from the encoded information of each element,  which will be explained in the following content.
In our current study, a three stage and third order explicit Runge--Kutta method is used for time integration~\cite{jiang1996efficient}. From time step $t^n$ to time step $t^{n+1}$, the solution update is  organized as 

\begin{equation}\label{eq:rk3}	
\begin{cases}
	u^1 = u^n + \Delta t R(u^n),\\
u^2 = \frac{3}{4} u^n + \frac{1}{4}u^1 + \frac{1}{4}\Delta t R(u^1),\\
u^{n+1} = \frac{1}{3} u ^n + \frac{2}{3} u^2 + \frac{2}{3} \Delta t R(u^2),\\
\end{cases}
\end{equation}
where $R =- \frac{\partial f}{\partial x}- \frac{\partial g}{\partial y}- \frac{\partial h}{\partial z}$, which is  referred to as the right hand side. We have attempted to migrate $p$-multigrid preconditioned implicit time integration methods developed for CPU hardware in~\cite{wang2020comparison, wang2022nonlinear} to GPU platforms with no satisfying results at the time of writing. The ease of building up a hierarchy of meshes for $h$-multigrid methods by using efficient octree algorithms is yet to be explored.
Throughout this paper, we are going to use ``$\dim$'' to indicate the dimension of the problem.

\subsection{Linear trees and Morton IDs}
\begin{figure}	
\begin{subfigure}[b]{0.49\textwidth}
\includegraphics[width=\textwidth]{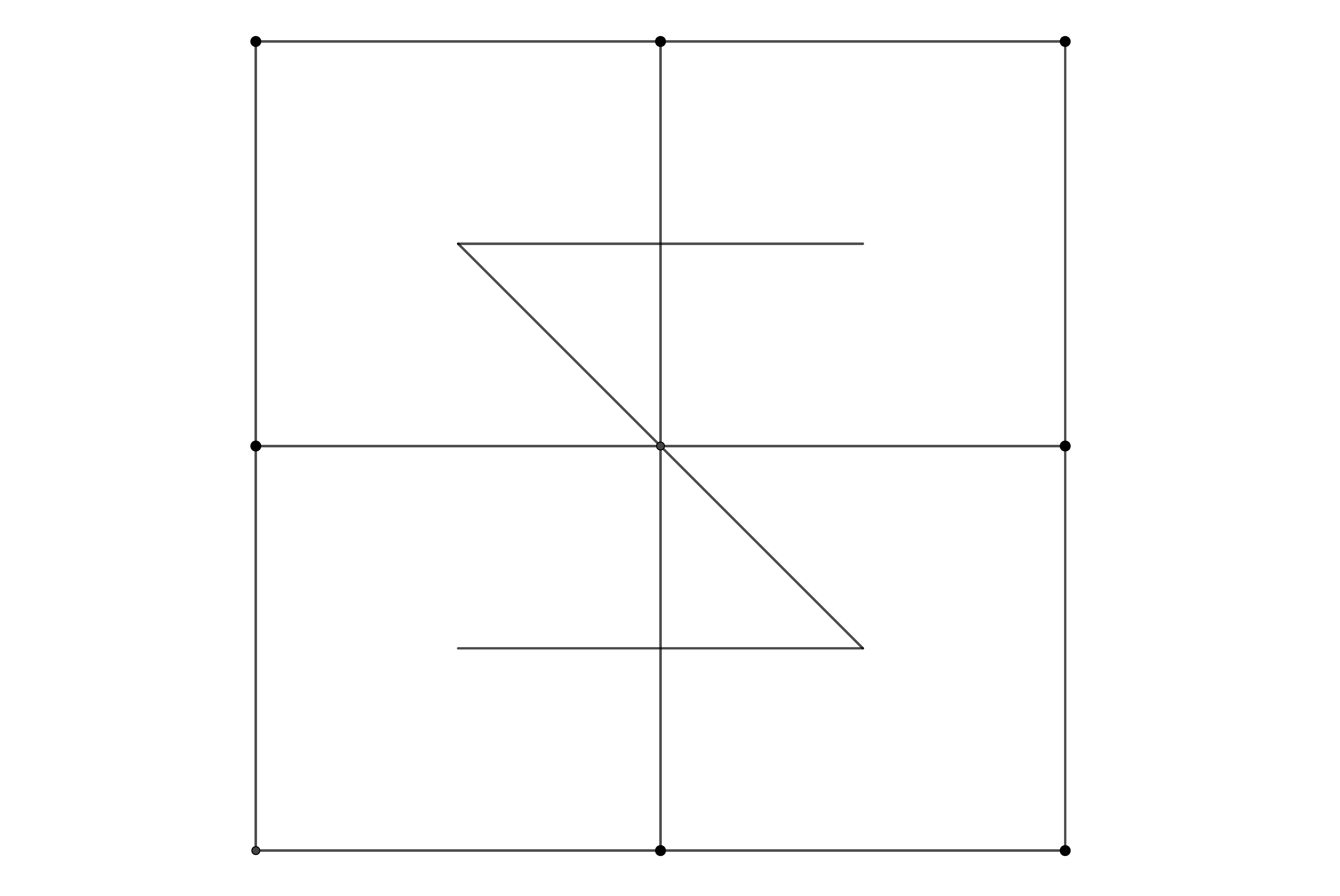} 
\caption{Z-shape filling curve}
\label{subfig:z-shape}
\end{subfigure}
\hfill
\begin{subfigure}[b]{0.49\textwidth}
\includegraphics[width=\textwidth]{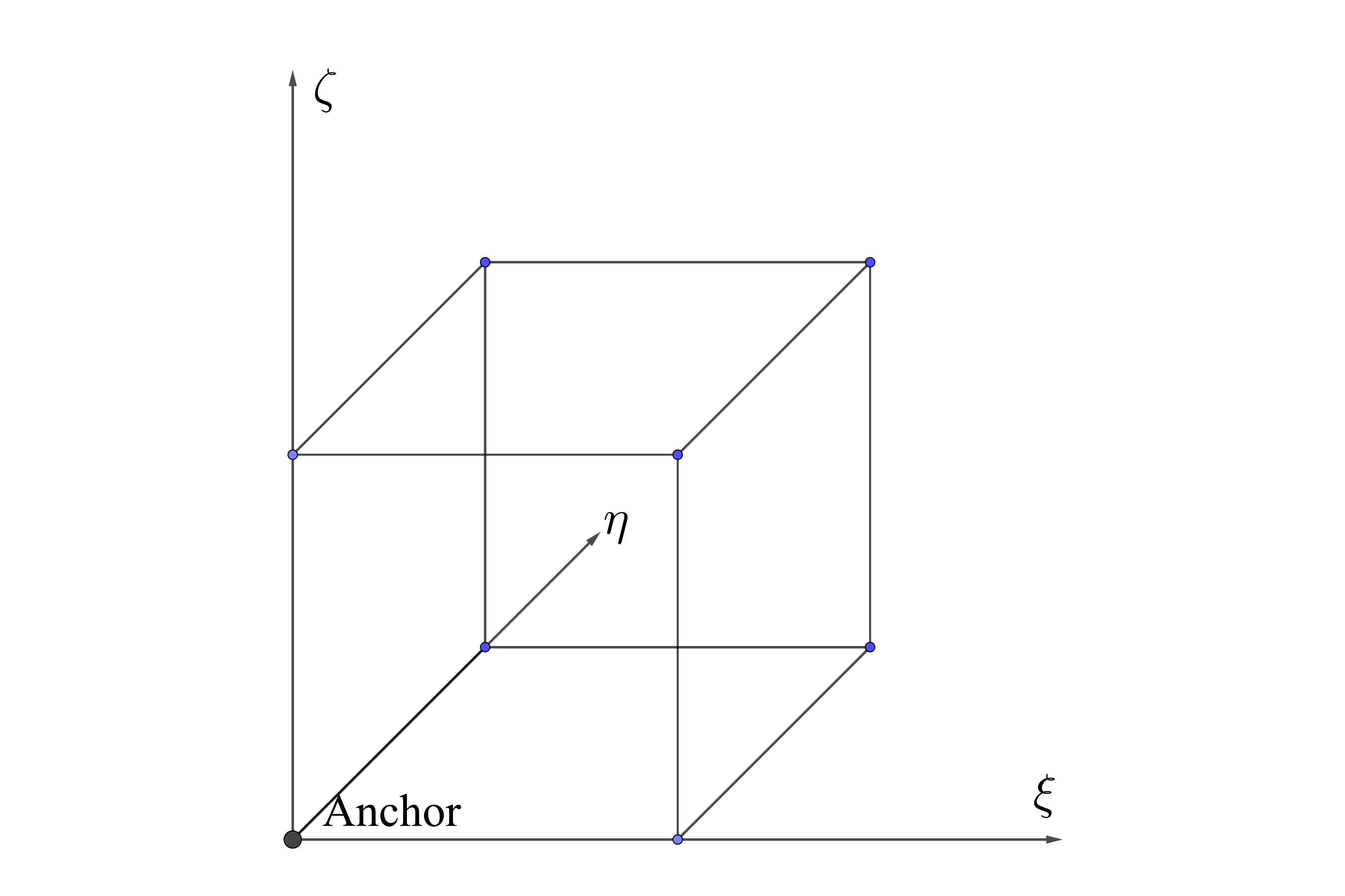} 
\caption{Anchor of an octant}
\label{subfig:anchor}
\end{subfigure}
\caption{Illustration of anchor and Z-curve.}
\label{fig:anchor_zshape}
\end{figure}

In tree structures, such as the binary searching tree, a leaf node is a node in the tree that does not have any child nodes. 
A linear tree (quadtree for 2D and octree for 3D) refers to the fact that only the leaf nodes of the tree are stored as a sorted array of integers where each integer represents a leaf node in the tree.  
A node is usually referred to as an octant in an octree. And we use an octant to represent a node in a quadtree for consistency. In this paper, we are concerned with using Morton encoding to get the Morton ID of each node of the tree.
The sorted array of Morton IDs is one type of space filling curves (SFC) and when it is visualized, we can observe zigzag shape of curves filling the entire domain, as shown in Figure~\ref{subfig:z-shape}, and they are usually called Z-shape SFC.

Morton encoding can encrypt both the coordinates and the depth of an octant in the tree. With predefined spatial resolution, one would take the coordinates of one corner of the octant and convert them into an integer. The encoded corner is referred to as the anchor of the octant~\cite{sundar2008bottom}.
As shown in Figure~\ref{subfig:anchor}, in this paper we use the left-front-bottom corner as the anchor.
In Morton encoding, the inclusive region of the octant $O_j$ at depth $D_j$ is defined as $[\mathbf{x}_j, \mathbf{x}_j + \Delta x_j )$, where $\mathbf{x}_j$ is the coordinates of the anchor and $\Delta x_j$ is the width of the octant.
In this paper, we only consider perfect cubes as the octree mesh, i.e., the widths of any octant in the tree in all directions are the same. As the size of the octants are sorted out at different depths, one can always convert the coordinates of the mesh from floats to integers. To distinguish, we use $[\mathbf{x}^f_j, \mathbf{x}^f_j + \Delta^fx_j )$ to denote the region of an octant in the actualy physical domain, and we use $[\mathbf{x}^I_j, \mathbf{x}^I_j + \Delta^Ix_j )$ for the normalized integer representation of the tree domain. 
For any octant $O_j$ at depth $D_j$, we can perform the Morton encoding by interleaving the bits of different components of $\mathbf{x}_j^I$ first and then append the level information. If we define the maximum depth allowed in the octree as $D_{max}$,
the overall algorithm is described in Algorithm~\ref{alg:morton_encoding}, following the work of Sundar el al.~\cite{sundar2008bottom}. Note the difference between raw Morton ID $\overline{O_j}$ and Morton ID $O_j$ is that the former only has the coordinate information.

\begin{algorithm}
\caption{Morton Encoding}\label{alg:morton_encoding}
\begin{algorithmic}[1]
	\State Determine the maximum tree depth $D_{max}$ and minimum element size $\Delta x_{min}^f$ for the application and calculate the maximum number of bits, $m_b$, that is needed to represent $D_{max}$.
	\State Convert the float coordinates $\mathbf{x}_j^f$ of any octant into integer coordinates $\mathbf{x}_j^I$ based on $\Delta x_{min}^f$ and determine the depth of $D_j$.
	\State Interleave the bits of three components of the coordinate as $  \boldsymbol{x}^{D_j-1}  \ldots \boldsymbol{x}^1   \boldsymbol{x}^0$ to get the raw Morton ID $\overline{O_j}$,  where $\boldsymbol{x}^{k} = x_3^k x_2^k x_1^k$ for 3D and $\boldsymbol{x}^{k} = x_2^k x_1^k$ for 2D in this algorithm, which means retrieving the $k$-th bit of each component of $\mathbf{x}_j^I$ and putting them together to get a binary number of 2/3 bits for a 2/3-dimensional problem.
	\State Add padding bits to the raw Morton ID $\overline{O_j}$ as $  \boldsymbol{x}^{D_j-1}  \ldots \boldsymbol{x}^1   \boldsymbol{x}^0 \underbrace{0}_{\text{repeat}\ \dim \cdot (D_{max} - D_{j})}$.
	\State Append the binary representation of the depth $(D_j)_2$ to the result obtained in previous step as 
	$  \boldsymbol{x}^{D_j-1}  \ldots \boldsymbol{x}^1   \boldsymbol{x}^0 \underbrace{0}_{\text{repeat}\ \dim \cdot ( D_{max} - D_{j})} \underbrace{(D_j)_2}_{m_b \text{bits}}$ to get the Morton ID $O_j$ which includes all the information that we need.
\end{algorithmic}
\end{algorithm}
Once we obtains the Morton IDs of all octants in the tree, we sort the linear tree in ascending order to get $\mathcal{T}$, which is the post-order traversal of the tree.
We refer readers to the work of Sundar el al.~\cite{sundar2008bottom} for other interesting properties. Given the Morton ID of an octant, one can always decode it to retrieve the normalized integer coordinates as well as the width of the octant. Therefore, there is no need to save the non-uniform element size information explicitly.
There are various ways of efficient implementation of interleaving the integer bits (Step 3 in Algorithm~\ref{alg:morton_encoding}). We recommend the open-source software \verb*|libmorton|~\cite{libmorton18} for a collection of different approaches. In our GPU implementation, we utilize the magic number approach.

\section{GPU-based tree algorithms for discontinuous methods}\label{sec:tree_manipulations}
Without some efficient thread-safe data structures, especially hash tables, on GPU, we have to redesign the algorithms for tree manipulations. When we combine linear trees with adaptive CFD methods, efficient algorithms of building up the mesh connectivity will be equally important as tree completion and balancing algorithms. We are going to articulate the algorithms that we design for efficient tree manipulations and connectivity query. Some common algorithms such as sorting, binary search, etc., are from the open source library THRUST from NVIDIA~\cite{bell2012thrust}. Detailed coding techniques will be explained if it is important for readers to understand how the algorithms work. 
\subsection{Tree construction}
A set of octants is said to be complete if the union of these octants covers
the entire domain. Alternatively, one can also define a complete linear tree as a linear tree in which every interior node has exactly eight child nodes in 3D and four child nodes in 2D~\cite{sundar2008bottom}. The algorithm that we use for tree completion follows the idea developed by Sundar et al.~\cite{sundar2008bottom} with novel optimizations for the GPU hardware. 
\begin{algorithm}
\caption{Tree construction on GPU hardware using shared memory and atomic counters}\label{alg:contruction}
\begin{algorithmic}[1]
	\State Encoding the scattered points $\mathbf{x}_j$ into Morton ID $O_j$. And sort the array in ascending order to get the initial incomplete tree $\mathcal{T}_{init}$. 
	\State For two consecutive octants $O_j$ and $O_{j+1}$ of $\mathcal{T}_{init}$, find the largest common ancestor (LCA). 
		\begin{itemize}
			\item Split the LCA to get $2^{\dim}$ child octants $C_i$, where $i = 1,\ldots,2^{\dim}$.
			\item Update the parent octants of $O_j$ and $O_{j+1}$ from $C_i$;
			\item For the rest octants in $C_i$, store any $O_j<C_i<O_{j+1}$ in the shared memory for new block-local array of octants $\mathcal{T}_{loc}$.
			\item Repeat the procedure separately for  parents of $O_j$ and $O_{j+1}$, until $O_j$ and $O_{j+1}$ can be found in the new array of $C_i$.
		\end{itemize}
	\State Sort the block-local array $\mathcal{T}_{loc}$ and remove the duplicates within the kernel function. In the meantime, use block-local atomic counter to record the number of new octants in each block of threads.
	\State Update global atomic counter with the local one and write the block-local array to tail of the global array $\mathcal{T}_{init}$.
	\State Sort the tree in ascending order using \verb|thrust::sort|.
	\State Linearize the tree to get the complete tree $\mathcal{T}$.
\end{algorithmic}
\end{algorithm}
An efficient GPU implementation will need to make use of the shared memory within one block of threads to store the newly generated octants and then write the results into global memory of the GPU card. Two atomic counters are used in our implementation. The first one is a block-local counter to record locally how many new octants are created and the second one is a global counter to record and synchronize among blocks for all the new octants.
One caveat is that since the shared memory is used to store the newly generated Morton IDs, the total amount could exceed the full capacity of the shared memory under rare circumstances.  Out of shared memory in tree algorithms is not as intimidating as it sounds. It just means the sparsity of the initial octants cause excessive octants generation within one GPU block. We can always repeat the procedure when out of shared memory happens and note that for next round, the problem will be greatly alleviated since we have already injected a significant amount new octants into the tree.
We need to linearize the tree, which refers to removing the duplicates and overlaps in the tree.

\textit{Linearized tree}. A linearized tree is a non-overlapping and complete tree. We propose the following simple algorithm for tree linearization as documented in Algorithm~\ref{alg:linearization}. The designed algorithm fits data parallel paradigm very well by reversing the tree and utilizing the \verb|thrust::unique| algorithm in THRUST.
\begin{algorithm}
\caption{Tree linearization using THRUST}\label{alg:linearization}
\begin{algorithmic}[1]
\State Reverse the tree (change the tree from ascending order to descending order).
\State Perform the algorithm  \verb|thrust::unique| in the thrust library with a binary predicate which tells if two consecutive Morton IDs are overlapping.
\State Reverse the remaining tree back.
\end{algorithmic}
\end{algorithm}
\subsection{2:1 balance of the linear tree}
The linearized complete tree from Algorithm~\ref{alg:contruction} is not a 2:1 balanced tree. 
2:1 balance refers to the depth difference of any two adjacent octants must not exceed 2. In this paper, we pursue strong 2:1 balance, namely, any two adjacent octants that share either an edge or a face must not have tree-depth difference larger than 1. Similar to polynomial balancing in polynomial adaptation~\cite{wang2020dynamically}, the canonical method of balancing the tree is to start from the finest resolution to ensure the 2:1 rule is enforced and then repeat the same process for coarser resolution one by one. The most challenge part of balancing a linear tree is to find the neighbors of any given octant in the tree. 
With a linear sorted tree represented by Morton IDs, the problem of finding neighbors of an octant is equivalent to finding the lower bound of the corresponding Morton ID in the sorted array. 

\begin{figure}
\centering
\includegraphics[width=0.8\textwidth]{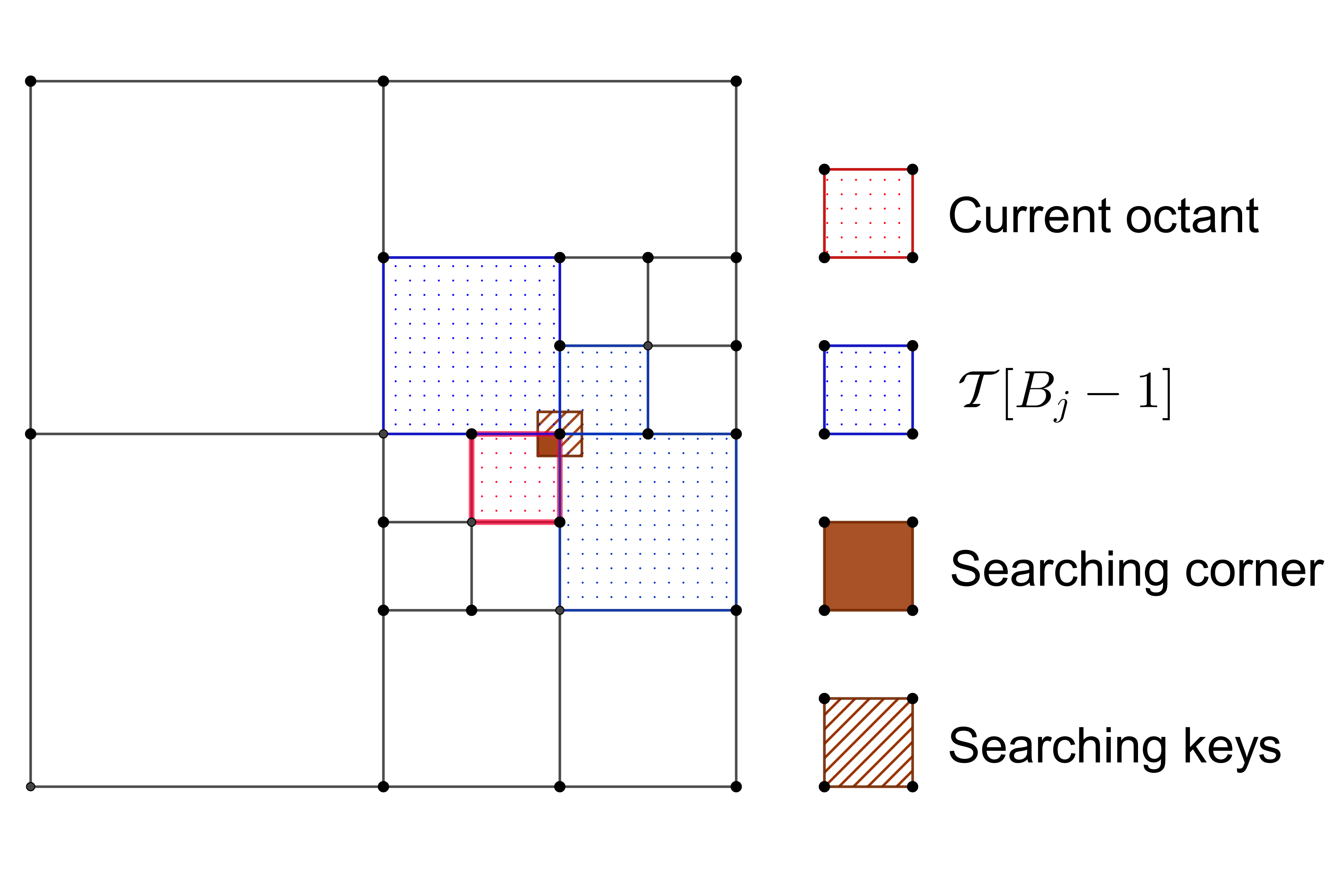}
\caption{Illustration of the searching keys for neighbors of an octant.}
\label{fig:octant_searching}
\end{figure}

\textit{Lower bound}. For any given Morton ID $O_j$, we define the lower bound location of $O_j$ in the linear tree $\mathcal{T}$ as the location $\mathcal{B}_j$ where $\mathcal{T}[\mathcal{B}_j]\geq O_j$. We use notation $a[i]$ to indicate $i$-th entry in array $a$ in this manuscript. The algorithm \verb|thrust::lower_bound| in THRUST is used since it can efficiently find lower bounds of an input array in another array.
Based on the properties of Morton IDs, we have three significant scenarios, which will be used later, listed as follows:
\begin{equation}\label{lower_bound}
\begin{cases}
 \mathcal{T}[\mathcal{B}_j] = O_j, \text{ i.e., lower bound of $O_j$ in the tree is $O_j$ itself};\\
 \mathcal{T}[\mathcal{B}_j-1]  \ni O_j, \text{ i.e., the octant on the left of the lower bound of $O_j$ is the parent of $O_j$};\\
 \mathcal{T}[\mathcal{B}_j] - O_j = 1\ \text{and}\ \mathcal{T}[\mathcal{B}_j-1]  \in O_j, \text{i.e., the lower bound is the first child of octant $O_j$}.
\end{cases}
\end{equation} 
Note that in this paper, we use $A\in B$ to denote that $A$ is a child of $B$ and $B\ni A$ to denote $B$ is a parent of $A$.
We adopt the concept of searching corners used by Sundar et al.~\cite{sundar2008bottom}. A searching corner is the octant on the finest resolution that shares a vertex with both the current octant and the parent octant. The searching keys are octants that make the shared vertex as the center of a cube/square formed by the searching corner and keys. We remark that a searching key does not have to be in the tree and one can always find either the searching key itself or an ancestor of the searching key as long as the tree is complete and linearized. The idea of searching corner and corresponding searching keys are presented in Figure~\ref{fig:octant_searching} and we also visually show what is a lower bound for a given searching key.

The algorithm that we develop to balance tree at depth $D$ is presented in Algorithm~\ref{alg:balance}. The optimization here for multi-threading, compared to the work of Sundar et al.~\cite{sundar2008bottom}, is that we collect all the searching keys together and perform data parallel computing.
\begin{algorithm}
\caption{Balancing tree at depth $D$ on GPU}\label{alg:balance}
\begin{algorithmic}[1]
\State Collect all the Morton IDs that are on depth $D$ in the tree $\mathcal{T}$ and store them as $\mathcal{R}_{D}$.
\State Find all the searching keys of all the Morton IDs in $\mathcal{R}_{D}$ and store the unique values as $\mathcal{S}_{D}$.
\State For every searching key in $\mathcal{S}_{D}$, find the location of its lower bound in the tree $\mathcal{T}$ and store the results in $\mathcal{B}_{D}$.
\State  For each searching corner $\mathcal{S}_j$ in $\mathcal{S}_D$, if  $\mathcal{T}[\mathcal{B}_{D,j}-1] $ is a parent of $\mathcal{S}_j$, refine $\mathcal{T}[{\mathcal{B}_{D,j-1}}] $ and store any octant which is not a parent of $\mathcal{S}_j$ in the shared memory. Repeat the process for the new parent octant of $S_j$ until the depth of these new octants reaches $D+1$. 

\State Perform a block-lock removal of duplicates and use a block-local atomic counter to record the total unique Morton IDs per block.
\State Use the block-local counter to update the global atomic counter of the total number of new octants and in the meantime write new octants to the tail of tree $\mathcal{T}$.
\State Sort and linearize the updated tree $\mathcal{T}$.
\end{algorithmic}
\end{algorithm}
In the course of balancing a tree, the searching keys and local refinement of target octants will end up with lots of duplicated child octants. 
One could utilize a hash table to quickly remove duplicates on the fly as Step 4 of Algorithm~\ref{alg:balance} generates new octants if the algorithm were implemented for CPU hardware.  On GPU hardware, no efficient thread-safe hash tables is available to fulfill our goal of removing duplicates.  In terms of removing the duplicates, we use the same approach as in Algorithm~\ref{alg:contruction}. Note that as long as the tree is completed, we do not have to perform tree completion anymore in the 2:1 balancing algorithm. One additional note is that in numerical experiments, out of shared memory is more likely to happen in Algorithm~\ref{alg:balance} instead of Algorithm~\ref{alg:contruction} in the phase of building an initial mesh. In CFD simulations, since the mesh before adaptation is already completed and balanced, it would not be an issue.

\subsection{Interface data structure and connectivity}\label{sec:face_connectivity}
A linear tree of octants is not equivalent to a mesh of elements. For methods like continuous Galerkin methods, nodal information such as how a node is connected to elements is needed. Finding such information is trivial on CPU hardware with hash tables but challenging if done entirely on GPU. Fortunately, for discontinuous method, such as the flux reconstruction method used in the work, we only needs to know how elements communicate through faces.

\begin{figure}
 \centering
 \begin{subfigure}{0.45\textwidth}
 \includegraphics[width=\textwidth]{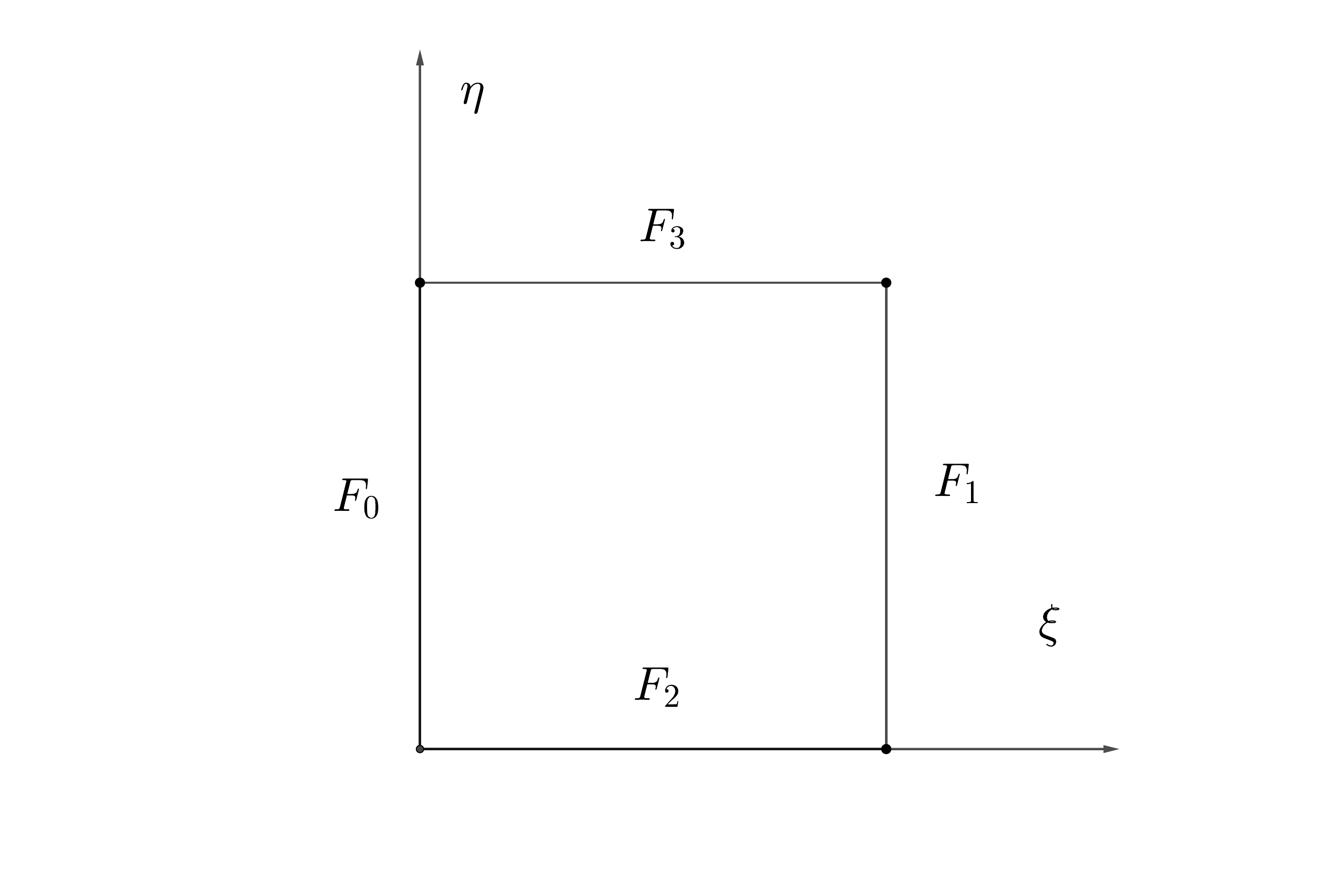}
 \caption{Faces in a 2D element}
 \label{subfig:face2d}
 \end{subfigure}
\begin{subfigure}{0.45\textwidth}
\includegraphics[width=\textwidth]{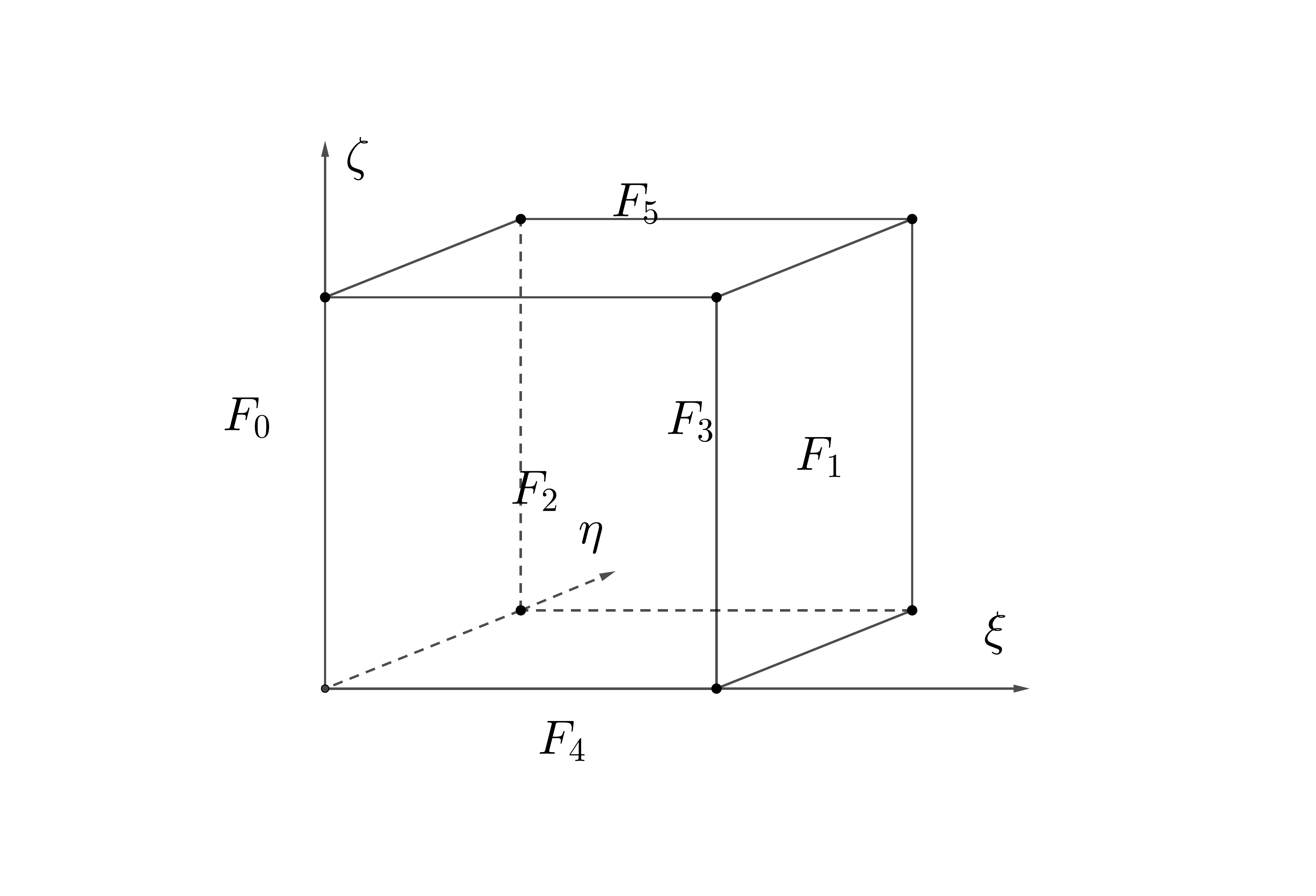}
\caption{Faces in a 3D element}
\label{subfig:face3d}
\end{subfigure}
 \caption{Face numbering for 2D and 3D elements.}
 \label{fig:face_numbering}
 \end{figure}

\textit{Data structures.}
For a tree of length $M$, we define an array $\mathcal{F}$ of length $n_f\cdot M$, where $n_f =2\cdot\dim$ is the number of faces that an octant has and $\dim$ is the dimension of the problem. Following \verb|C++| convection of array indexing, the entries in $[n_f\cdot \iota_e, n_f\cdot (\iota_e+1) )$ stores the indices of the neighboring elements of the current element $\iota_e$. For any octant, we define the face indices as illustrated in Figure~\ref{fig:face_numbering}.

\textit{Face matching.} With the linear data structure defined for face matching, the remaining question is to find 
the neighbors. Similarly, we need to identify the searching keys first. The nuance here is that we define the searching keys on the same depth as the current octant. There are three types of faces in the nonuniform mesh represented by a linear tree, namely, conforming faces, nonconforming faces and single-sided faces. Both conforming and nonconforming faces are interior faces and single-sided faces will only appear on the boundary. 
The algorithm to find the matching pair for the faces that an octant have is organized in Algorithm~\ref{alg:face_con}.
 
\begin{figure}
	\centering
	\includegraphics[width=0.8\textwidth]{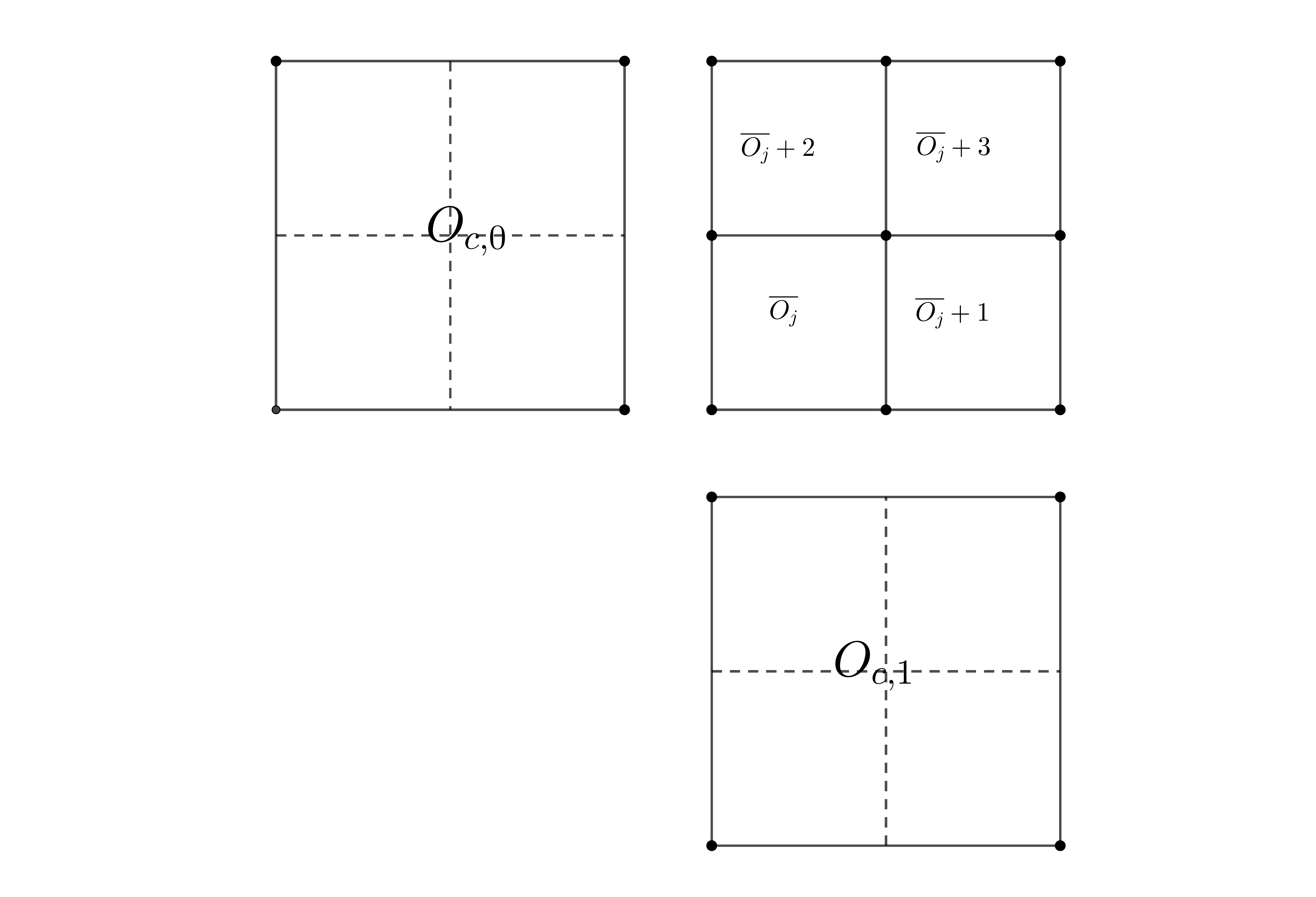}
	\caption{Illustration of face matching in quadtree, where all siblings of $\overline{O}_j$ are in the tree.}
	\label{fig:2d_face_query}
\end{figure}

\begin{algorithm}
\caption{Build face connectivity}\label{alg:face_con}
\begin{algorithmic}[1]
	\State Find the search keys of any octant $O_j$ in tree $\mathcal{T}$ by calculating the Morton IDs of all its neighbors which share faces with it as shown in Figure~\ref{fig:interface_searching} and store in $S$.
	\State Find lower bounds in the current tree $\mathcal{T}$ for all keys and store in $B$.
		\begin{itemize}
		\item If $\mathcal{T}[B_{j}] = S_j$ , then the neighbor of $O_j$, namely, $S_j$ is in the tree and the  corresponding face of $S_j$ is the matching conforming face.
		\item If $\mathcal{T}[B_j-1] \ni S_j$, then the neighbor of $O_j$ is the parent of $S_j$. The corresponding face of $\mathcal{T}[B_j-1]$ is the coarse face of a nonconforming pair.
		\end{itemize}
		
\end{algorithmic}
\end{algorithm}
When we build up the searching keys $S$ in Algorithm~\ref{alg:face_con}, these keys do not have to be in the tree, similar to the searching keys for 2:1 balance. 
In Step 2 of Algorithm~\ref{alg:face_con}, the spatial relationship of two sides of one conforming pair is straightforward.
For the $i_f$-th face of $O_j$, its matching conforming face will be the $(i_f + 1)$-th face of the neighboring element if $i_f$ is an even number or $i_f-1$ if $i_f$ is an odd number, following the numbering convention shown in Figure~\ref{fig:face_numbering}.  
Note that raw Morton ID $\overline{O_j}$ is obtained by removing the padding zero bits and depth bits of $O_j$. 
Taking Face 0 of $\overline{O}_j$ and $\overline{O}_{j+2}$ in Figure~\ref{fig:2d_face_query} as an example, they both have Face 1 of $O_{c,0}$ as the matching face. When we need to project the working variables from the fine faces (Face 0 of $\overline{O}_j$ and $\overline{O}_{j+2}$) to the coarse face (Face 1 of $O_{c,1}$), we can retrieve the last two bits of $\overline{O}_j$ and $\overline{O}_{j+2}$ to get $0$ and $2$ in decimal representation. As we define the first contributor along one axis will be the one that has smaller coordinates, we will know $\overline{O}_j$ is the first contributor and $\overline{O}_j+2$ is the second in the projection procedure of common flux calculations.  Similarly, for Face 2 of $\overline{O}_j$ and $\overline{O}_j +1$, the last two bits of these two numbers will be $0$ and $1$ and $\overline{O}_j$ is first contributor for the nonconforming face formed with coarser octant $O_{c,1}$. We remark that for any $2^{\dim -1}$ siblings that form a nonconforming face with another coarser element, we do not care if the other  $2^{\dim -1}-1$ siblings are interior nodes or leaf nodes in the tree.
 
\begin{figure}
	\centering
	\includegraphics[width=0.8\textwidth]{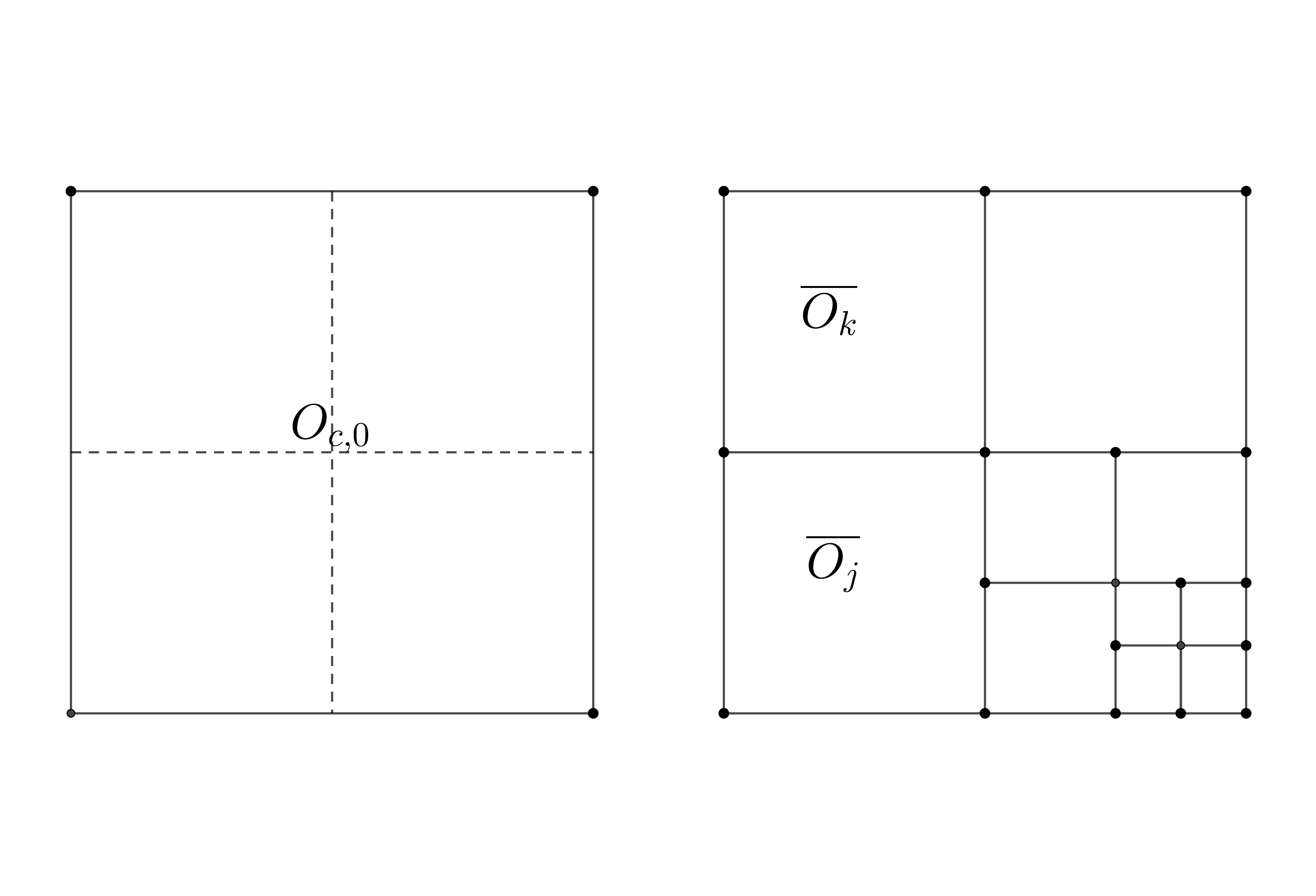}
	\caption{Illustration of face matching in quadtree, where $\overline{O}_k$ and $\overline{O}_j$ cannot be inferred from each other.}
	\label{fig:2d_face_query_bad_eg}
\end{figure}

Readers would be interested in why we do not try to find the neighbors of finer resolution for a coarser elements instead. One example is presented in Figure~\ref{fig:2d_face_query_bad_eg}, for $O_{c,0}$, one can find either $\overline{O}_k$ or $\overline{O_j}$; however, they cannot be inferred from each other. That means, in order to make things work, we will need to first check if $O_{c,0}$ on this side has one or tow neighboring elements and we also need to store the contributors in order.  This is definitely inefficient compared to the approach that we are using.

More information regarding how the face connectivity information will be used in discontinuous Galerkin methods will be illuminated in next section. It is noteworthy that what is needed for discontinuous finite Galekin methods can simply be obtained by performing series of binary search for the search keys. The methods are indifferent to nodal connectivity, which would normally require data structures like hash tables to efficiently obtain. The attributes that discontinuous methods have make them better suited for AMR on GPU than continuous Galerkin methods.

\begin{figure}
	\centering
	\includegraphics[width=0.8\textwidth]{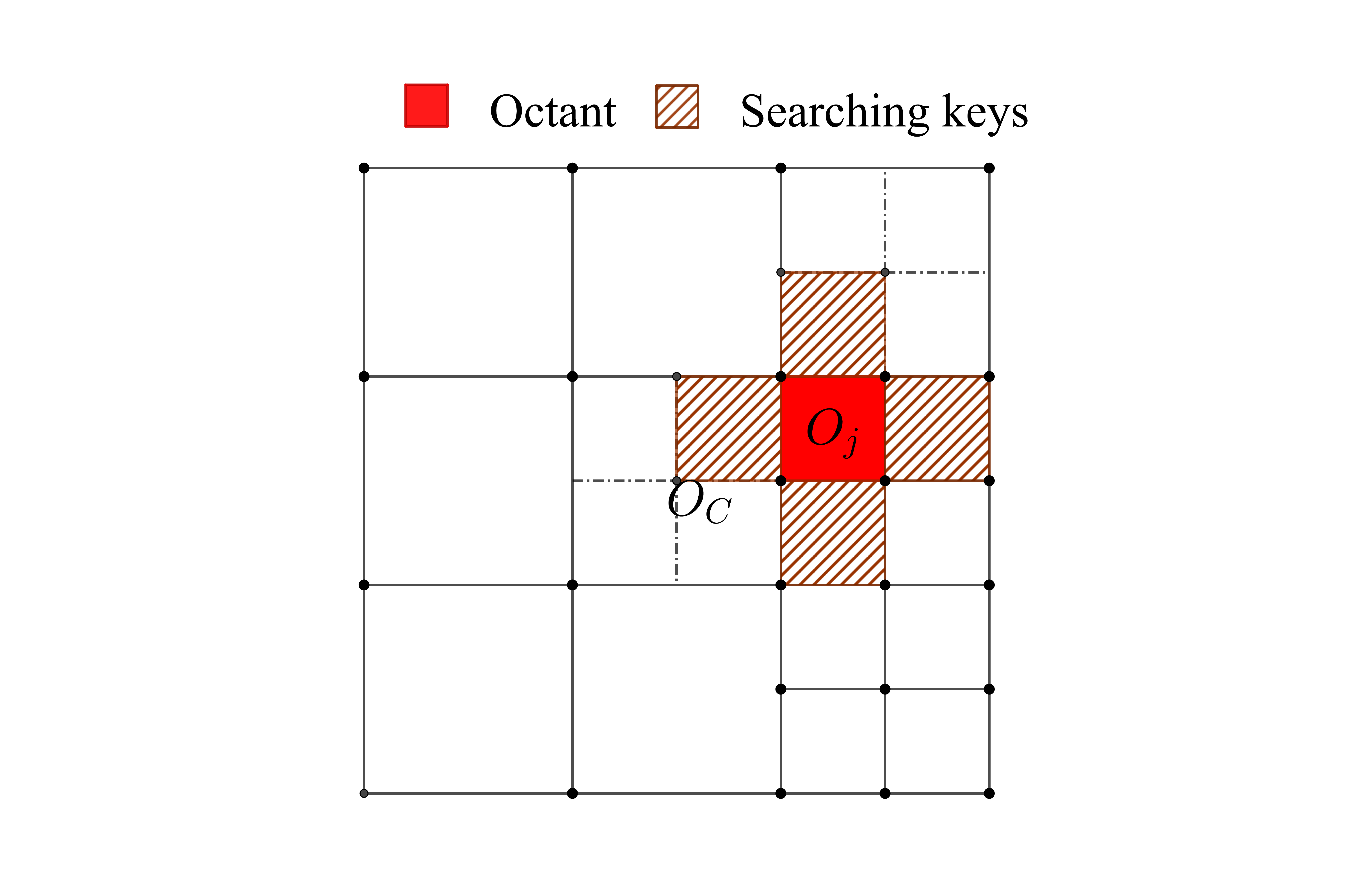}
	\caption{Illustration of the searching keys of an octant for interface matching.}
	\label{fig:interface_searching}
\end{figure}

\section{The $h$-adaptive flux reconstruction method}\label{sec:adaptive_fr}

\subsection{Adaptation procedure}
The procedure that we use to carry out adaptation is organized as in Algorithm~\ref{algo:adapt_procedure}.
For coarsening, one has to first rectify the marking in Step 3 of Algorithm~\ref{algo:adapt_procedure} simply due to the fact that an interior node of the tree could have child nodes and grandchild nodes at the same time. Moreover, we would have to identify those interior node in Step 4 and then remove the child nodes in Step 5 while no extract action is needed for refinement. The tree linearization will automatically remove any octant which is refined in the array. Our algorithm of adaptation procedure also implies that we can at most refine or coarsening an octant once in one adaptation procedure. 
Step 4 relies on the properties of the siblings illustrated in Figure~\ref{fig:2d_face_query}, where in the sorted array $2^{\dim}$ siblings are consecutively stored with increment of the raw Morton ID equals to 1.

Adaptation criteria for CFD simulations of unsteady problems have been widely studied in literature. There are featured-based adaptation indicators~\cite{wang2020dynamically}, discretization-error based, truncation-error based, adjoint-based adaptation based~\cite{fidkowski2010entropy}, etc. Discussion or evaluation of these different methodologies will beyond the scope of this work and we recommend the work of Naddei et al.~\cite{NADDEI2019508} and Ims and Wang~\cite{ims2022comparison}. When AMR is coupled with the flux reconstruction method, we would like to, in this work, focus on the performance of tree balancing, connectivity query, data transfer, compared to that of the flux reconstruction solver. Therefore, we are going to use a simple approach based on the analytical solution of a simple problem in the numerical experiment studies.

\begin{algorithm}
	\caption{Adaptation procedure of the tree}\label{algo:adapt_procedure}
	\begin{algorithmic}[1]
		\State Evaluate the adaptation flags for each element in the mesh. Elements will be marked with flag -1 and 1 for coarsening and refinement, respectively. Flag 0 means the element will remain the same.
		\State Refinement of any octant will be honored via directly splitting the octant and append the child octants to the tail of tree $\mathcal{T}$. Similar techniques of using block-local and global atomic counters are used as that in 2:1 balance.
		\State Rectify coarsening flags by reverting flag -1 to 0 if any octant does not have all its siblings existing in $\mathcal{T}$.
		\State If all child octants of an interior node are marked for coarsening, collect the parent Morton ID in $\mathcal{P}$. 
		\State  Remove those Morton IDs in $\mathcal{T}$ which can find its parent in $\mathcal{P} $ and append $\mathcal{P}$ to $\mathcal{T}$ afterwards. 
		\State Sort and linearize $\mathcal{T}$.
	\end{algorithmic}
\end{algorithm}

\subsection{Solution data transfer}
Data transfer between the new tree $\mathcal{T}_{n}$ and old tree $\mathcal{T}_{o}$ on GPU relies on the mapping between the two trees.
There are three types of data transfer in an adaptation procedure.

\textit{Direct copy}. Octants in $\mathcal{T}_n$ and $\mathcal{T}_o$ that have the same Morton IDs are elements which are kept as the same in one adaptation procedure. A direct copy of the solution variables is sufficient. 
 
\textit{Interpolation}. When an octant is split into $2^{\dim}$ children, the working variables in these children can be interpolated from the parent octant. If the anchor of any octant has coordinates $\mathbf{x}_0$, when the octant is refined, the anchor of the $k$-th children of the octant has the coordinates as 
$\mathbf{x}_0 + \mathbf{i}\cdot \Delta /2 $, where $\Delta$ is the width of the octant. And $\mathbf{i}$ is the local coordinate of the children octant and each component of $\mathbf{i}$ is either 0 or 1.
Let the solution in the parent octant $P$ denoted as 
\begin{equation}\label{eq:parent_soln}
    u^P(\mathbf{\xi}) = \sum_{i=1}^{N} l_i(\boldsymbol{\xi})u_i, 
\end{equation}
where $N = (k+1)^{\dim}$ is the  total number of Lagrange polynomials and $l_i(\boldsymbol{\xi})$ is the tensor product of the 1D Lagrange polynomials.
In the child octant, the solution on local solution points are interpolated from the parent octant using  Eq.~\eqref{eq:parent_soln}. Note that the interpolation coefficients can be directly evaluated within the kernel function. We opt to use interpolation instead of projection for refinement to reduce the copy of a dense projection matrix into the shared memory here. Moreover, as we do point-wise calculations for each solution point, direct evaluation of the interpolants on device can increase the number of elements being processed within one block of threads.

\textit{Projection}.  When $2^{\dim}$ siblings are merged into one parent octant, a $L_2$ projection is used to transfer the data from child octants to the parent octant. Given $K_\iota$, where $\iota=1,2,\ldots,2^{\dim}$, are the child octants which are fused into the parent octant $P$. Let $\boldsymbol{z}$ be the local coordinate of the child octant and the solution within the child octant can be organized as 
\begin{equation}\label{eq:child_soln}
    u_\iota^K(\boldsymbol{z}) = \sum_{i=1}^{N} \hat{l}_i (\boldsymbol{z})\hat{u}_{i}.
\end{equation}
    We would like to minimize the inner product error by requiring 
    \begin{equation}\label{eq:vol_l2_projection}
       \int_{\Omega_{K_\iota}} u^P(\boldsymbol{\xi})  l(\boldsymbol{\xi}) d\boldsymbol{\xi}= \sum_{\iota=1}^{2^{\dim} } \int_{\Omega_{K_\iota}}  u^K_\iota(\boldsymbol{z}(\boldsymbol{\xi})) l(\boldsymbol{\xi}) d\boldsymbol{\xi}.
    \end{equation}   
If we substitute corresponding terms with Eq.~\eqref{eq:parent_soln} and Eq.~\eqref{eq:child_soln}, we can organize Eq.~\eqref{eq:vol_l2_projection} as 

\begin{equation}\label{eq:l2_projection_components}
	\int_{\Omega_{P}} \left(\sum_{i=1}^N l_i(\boldsymbol{\xi}) u_i\right) l_j(\boldsymbol{\xi}) d\boldsymbol{\xi}= \sum_{\iota=1}^{2^{\dim} } \int_{\Omega_{K_\iota}}  \left(\sum_{i=1}^N \hat{l}_i(\boldsymbol{z}(\boldsymbol{\xi})) \hat{u}_i\right) l_j(\boldsymbol{\xi}) d\boldsymbol{\xi}.
\end{equation}  
After some manipulations of Eq.~\eqref{eq:l2_projection_components}, we can get the discrete formulation as
\begin{equation}
\boldsymbol{M} \boldsymbol{ u}^P = \sum_{\iota=1}^{2^{\dim} } \boldsymbol{S}_\iota \boldsymbol{u}^K_\iota,  
\end{equation}
where $\boldsymbol{M}$ and $\boldsymbol{S}_\iota$ are expressed as 
\begin{equation}
	M_{ji} = \int_{\Omega_{P}}  l_j l_i d \boldsymbol{\xi}\ \text{and}\ S_{\iota, ji} = \int_{\Omega_{P}} l_j \hat{l}_i d \boldsymbol{\xi}.
\end{equation}
All the projection operators that are described in this section are stored in the global memory of the GPU hardware.
At this point, readers should be familiar with the usage of lower bounds and would not be surprised that we are going to use a similar approach to find the mapping between old tree $\mathcal{T}_o$ and new tree $\mathcal{T}_n$. In Algorithm~\ref{algo:mapping_old_new_tree}, we present the algorithm to identify the mapping. 
The Step 5 of Algorithm~\ref{algo:mapping_old_new_tree} is needed because when we use atomic counters to write information into the data arrays on the global memory, there is not rule that regulates which thread writes first. By sorting the IDs, we can coalesce the memory access as much as possible in the data transfer procedure.
\begin{algorithm}
	\caption{Mapping between old and new tree}\label{algo:mapping_old_new_tree}
	\begin{algorithmic}[1]
		\State Find the lower bounds for all Morton IDs $O_j$ of the new tree $\mathcal{T}_{n}$ in the old tree $\mathcal{T}_{o}$ and store in $B$.
		\State If $\mathcal{T}_{o} [B_j] = O_j$, then this is a direct-copy octant. 
		\State If $\mathcal{T}_{o} [B_j-1]$ is a parent of $O_j$, then the $\mathcal{T}_{o} [B_j-1]$ is refined to get $O_j$.
		\State If $\mathcal{T}_{o} [B_j] - O_j = 1$,  then this octant $ \mathcal{T}_{o} [B_j]$ together with following $2^{\dim} -1 $ octants in the old tree are fused to get get $O_j$. 
		\State Store corresponding $B_j$ and $O_j$ for all different scenarios. After done, sort the $B_j$ and $O_j$ arrays based on the value of $O_j$.
	\end{algorithmic}
\end{algorithm}

\subsection{Common flux calculations on interfaces}
In previous section~\ref{sec:face_connectivity}, we present the data structures and algorithms to identify the face connectivity information that is needed for the flux reconstruction method. In this section, we are going to explain in detail how to perform the common flux calculations for inter-element communication in the flux reconstruction method. 
To avoid thread divergence, we will group the common flux calculations into three different types, namely, (1) conforming interior face pairs, (b) nonconforming interior face pairs, and (c) exterior boundary faces.
In this work, we will only consider periodic boundary conditions.  For periodic boundary faces, they eventually become interior faces. The special treatment is that when we try to identify the searching keys in the 2:1 balancing algorithm or the face connectivity query algorithm we need to take into account the periodicity of integer coordinates.
At a pair of conforming faces, one would only need to use the working variable from both sides $u_l$ and $u_r$ to calculate the common flux.
At nonconforming interfaces, we use the approach proposed by Kopriva~\cite{KOPRIVA1996475}.
The procedure to perform common flux calculations for mortar faces is described in Algorithm~\ref{algo:nonconforming}. 
 
\begin{algorithm}
	\caption{Procedure of common flux calculations}\label{algo:nonconforming}
	\begin{algorithmic}[1]
		\State Interpolate the solution from interior solution points to local flux points.
		\State Project the local solution on the flux points to the mortar face.
		\State Compute the common fluxes using a Riemann solver on the mortar face.
		\State  Project the common fluxes back to each side of the local elements.
	\end{algorithmic}
\end{algorithm}

\begin{figure}
	\centering
	\includegraphics[width=0.5\textwidth]{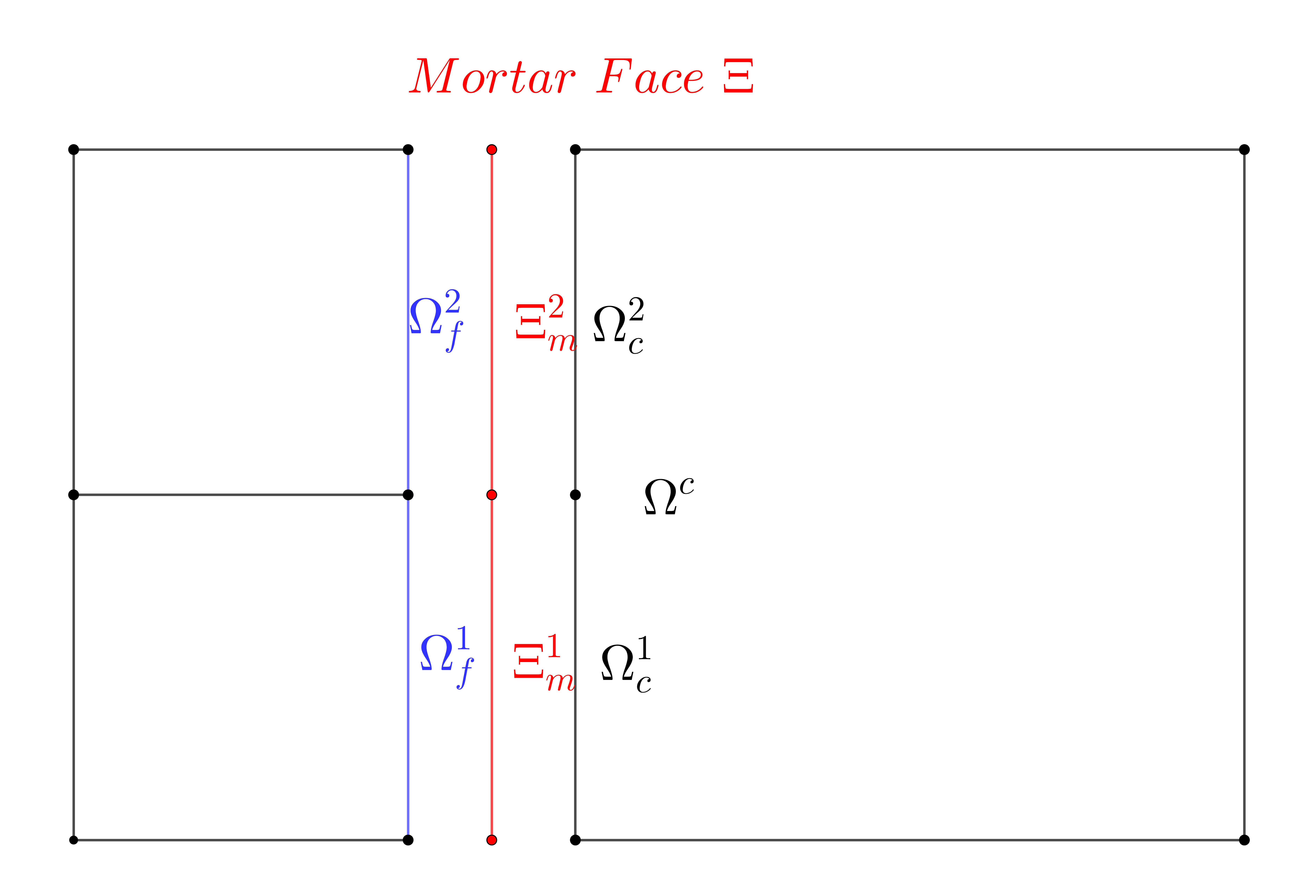}
	\caption{Illustration of the mortar interface.}
	\label{fig:mortar_face}
\end{figure}

Taking the illustration in Figure~\ref{fig:mortar_face} as an example,
the projection between the local faces $\Omega_f^{1,2}$ and mortar faces $\Xi^{1,2}_{m}$ is straightforward. Since they are conforming and the polynomial degrees are the same,  a direct copy is all we need.
The projection from local face $  \Omega_C$ to $\Xi_m^\iota$, $\iota =1,2,\ldots,2^{\dim-1}$ can be achieved by minimizing the inner product of the solution polynomials.
If we define the local solution polynomial on the interface $  \Omega_C$ as 
\begin{equation}
	u = \sum_{i=1}^{N} l_{i}(\boldsymbol{\xi}) u_{i},
\end{equation}
where $N=(k+1)^{\dim-1}$ for interfaces and the local solution polynomial on mortar faces $\Xi_m$ as
\begin{equation}
	\hat{u} = \sum_{i = 1}^{N} \hat{l}_{i}(\boldsymbol{\boldsymbol{z}}) \hat{u}_{i},
\end{equation} 
to compute discrete $\hat{\boldsymbol{u}}$ on each mortar face $\Xi_m^{\iota}$ , we seek a $L_2$ projection such that 

\begin{equation}
	\int_{\Xi_m^k} \left(\sum_{i=1}^{N} \hat{l}_i(\boldsymbol{z}) \hat{u}_{i} - \sum_{i=1}^{N} l_i(\boldsymbol{\xi}(\boldsymbol{z}))u_{i}\right) \hat{l}_j(\boldsymbol{z}) d \boldsymbol{z}= 0.
\end{equation}
Therefore, for each mortar face $\Xi_m^{\iota}$, one obtains
\begin{equation}\label{eq:interpolation_operator}
	\boldsymbol{M}_{\Xi_m^\iota}\hat{\boldsymbol{u}}_{\Xi_m^\iota} = \boldsymbol{S}_{\Xi_m^\iota}\boldsymbol{u}_{\Xi_m^\iota},\ M_{\Xi_m^\iota,ji} = \int_{\Xi_m^\iota} \hat{l}_j \hat{l}_i d \boldsymbol{z},\ \text{and}\ S_{\Xi_m^\iota,ji}=\int_{\Xi_m^\iota} \hat{l}_j l_i d \boldsymbol{z}.
\end{equation}
After the conserved variables are ready on the Mortar faces, we use a Rusanov solver for the common inviscid flux calculations in this work.
The projection of common fluxes from the mortar faces to the coarse face is a bit more complicated since $2^{{\dim}-1}$ mortar faces will contribute to the same coarse face. Similar to Eq.~\eqref{eq:vol_l2_projection}, 
we seek the flux polynomial that satisfy
\begin{equation}
\int_{\Omega_c} \left(F(\boldsymbol{\xi}) - \hat{F}(\boldsymbol{z}(\boldsymbol{\xi}))\right) l_j(\boldsymbol{\xi}) =0.
\end{equation}
We can rearrange the above equation as 
\begin{equation}
\int_{\Omega_c} \left(\sum_{i=1}^{N} l_i(\boldsymbol{\xi}) F_i\right) l_j(\boldsymbol{\xi}) d \boldsymbol{\xi}  = \sum_{\iota=1}^{2^{\dim-1}} \int_{\Omega_c^{\iota} } \left(\sum_{i=1}^{N} \hat{l}_i (\boldsymbol{z}_\iota(\boldsymbol{\xi})) \hat{F}_i^\iota\right) l_j(\boldsymbol{\xi})d \boldsymbol{\xi} 
\end{equation}
such that we can further simplify it into the discrete form as 
\begin{equation}\label{eq:projection_operator}
\boldsymbol{M} \boldsymbol{F} = \sum_{\iota=1}^{2^{\dim}-1} \boldsymbol{S}^\iota \hat{\boldsymbol{F}}^\iota, M_{ji} = \int_{\Omega_c} l_jl_i d \boldsymbol{\xi}, S^\iota_{ji} = \int_{\Omega^\iota_c} l_j \hat{l}_i d \boldsymbol{\xi}.
\end{equation}
The significance of minimizing the $L_2$ error of the common flux projection is to ensure the conservation of the numerical methods.
These projection operators in Eq.~\eqref{eq:interpolation_operator} and \eqref{eq:projection_operator} are pre-computed and stored in the global memory based on input polynomial degree.

In terms of implementation, one can directly load all necessary data into the kernel function and complete the projection in Eq.~\eqref{eq:interpolation_operator} within the same kernel function. However, for  the projection described in Eq.~\eqref{eq:projection_operator}, since multiple Mortar contributors could be scattered into different kernel functions, or even if they are in the same kernel function, they would write to the same destination as well. To protect the calculations from thread racing, we need to use the \verb*|atomicAdd| algorithm.

\section{Numerical results}\label{sec:numerical_exp}
In this section, a brief comparison of the GPU and CPU version of the tree completion and balancing algorithms is conducted first. We will focus on the performance of the adaptive solver in terms of both accuracy and computational cost are then analyzed. All simulations were run on a Windows Subsystem for Linux 2 (WSL2) with the Ubuntu 22.04.2 LTS Linux distribution while the host laptop has a NVIDIA  GeForce RTX 2060 GPU card. There are 1920 CUDA cores, 6GB memory for this GPU card and the bandwidth is 264.05 GB/s. The host laptop has an Intel i-10750H CPU with 6 cores.

\subsection{Tree completion and tree balancing}
In this section, we aim to test the efficiency of the tree completion and balancing algorithms for the algorithms that we developed. As a reference, we also implemented the CPU version of the code, which was made easy by the multi-platform support of THRUST and C++ template programming. The major difference is that in the CPU version we used \verb|std::unordered_map| from the standard template library (STL) in C++, which essentially is a hash table data structure and a \verb|std::mutex| is used to protect it from data racing. In all initial mesh generations of our numerical examples, we are going to use $D_{min}$ to generate a coarse background mesh and $D_{max}$ will be where the initial mesh points are encoded.

\begin{figure}
	\centering
	\begin{subfigure}{0.45\textwidth}
	\includegraphics[width=0.8\textwidth]{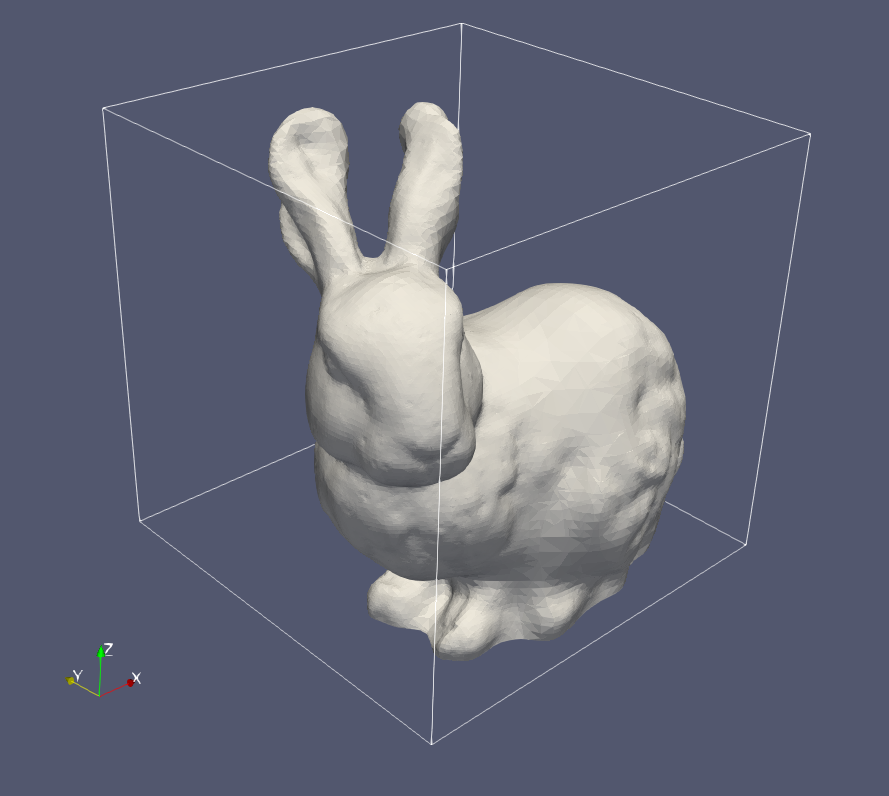}
        \caption{}
	\end{subfigure}	
    \hfill
	\begin{subfigure}{0.45\textwidth}
	\includegraphics[width=0.8\textwidth]{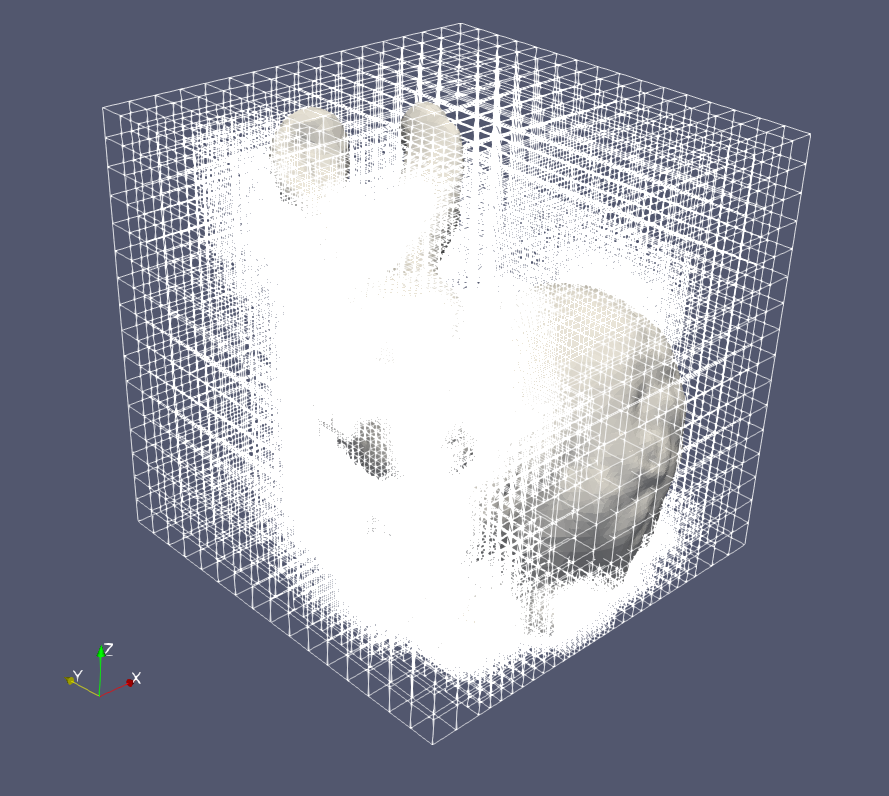}
        \caption{}
	\end{subfigure}	
	\caption{(a) Stanford bunny and (b) balanced tree.}
	\label{fig:bunny}
\end{figure}

\begin{figure}
	\centering
	\begin{subfigure}{0.45\textwidth}
	\begin{tikzpicture}	
		\begin{axis} [ybar = .05cm,
			bar width = 10pt,
			ymin=0,
			ymax=0.012,
			xmin=8,
			xmax=11,
			xlabel = $D_{max}$,
			ylabel = Runtime (seconds),
			xtick = {9,10},
			legend style={at={(0,1)},anchor=north west,nodes={scale=0.8, transform shape}}
			]
			
			\addplot [draw = black,
			line width = .4mm,
			pattern = {north west lines},
			pattern color = black
			] coordinates {(9, 0.0061103) (10, 0.0113826) };

			\addplot [draw = black,
			line width = .4mm,
			fill = black
			] coordinates {(9, 0.0065258) (10, 0.0069289)  };
	
			\legend {CPU, GPU};
		\end{axis}
	\end{tikzpicture}

	\caption{Tree completion}
	\end{subfigure}
	\hfill
	\begin{subfigure} {0.45\textwidth}
	\begin{tikzpicture}	
		\begin{axis} [ybar = .05cm,
			bar width = 10pt,
			ymin=0,
			ymax=0.9,
			xmin=8,
			xmax=11,
			xlabel = $D_{max}$,
			ylabel = Runtime (seconds),
			xtick = {9,10},
			legend style={at={(0,1)},anchor=north west,nodes={scale=0.8, transform shape}}
			]

			\addplot [draw = blue,
			semithick,
			pattern = {horizontal lines},
			pattern color = blue
			]   coordinates { (9,0.290557) (10,0.850036) };

			\addplot [draw = blue,
			semithick,
			fill = blue
			]   coordinates {(9,0.0568956) (10,0.2282)  };
			
			\legend {CPU, GPU};
		\end{axis}
	\end{tikzpicture}
	\caption{2:1 balancing}
	\end{subfigure}

	\caption{Computational cost of tree completion and 2:1 balancing for Stanford bunny.}
		\label{fig:bunny_cpu_gpu}
\end{figure}
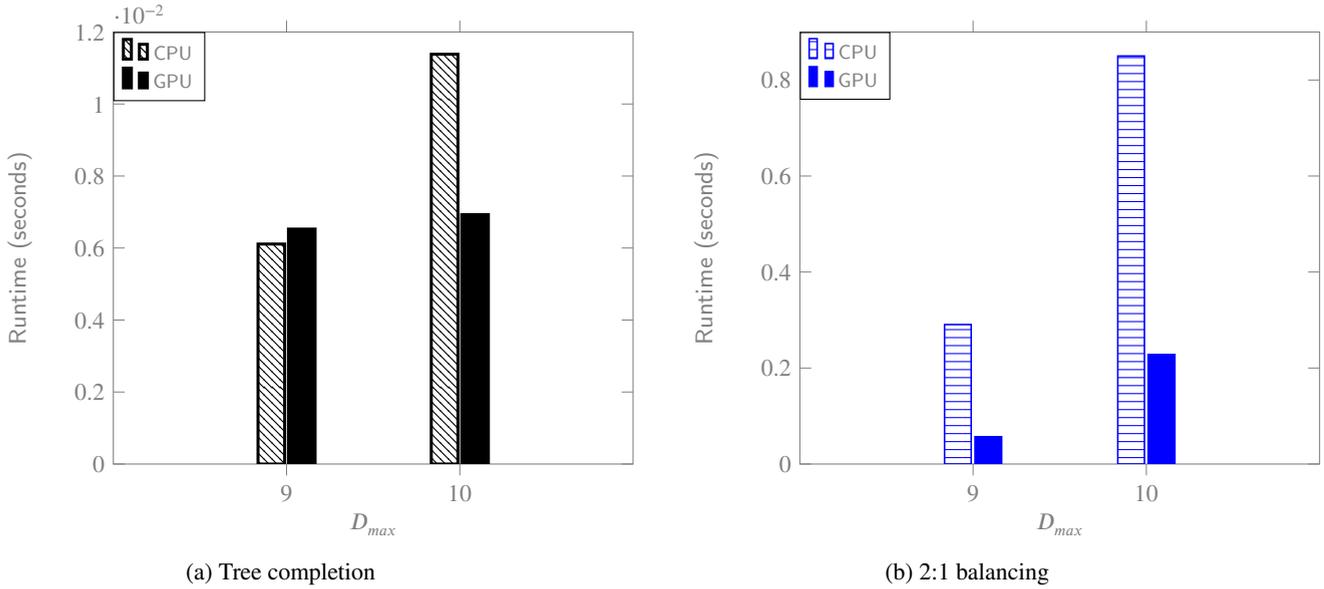

The Stanford bunny was used to test the robustness and performance of the developed tree completion and balancing algorithms. In Figure~\ref{fig:bunny}, we present the geometry and the balanced tree. 
For this testing case, the minimum tree depth $D_{min}$ was set as $D_{min} = 4$ for this case and the maximum tree depth $D_{max}=9,10$ were tested.  In Figure~\ref{fig:bunny_cpu_gpu}, the computational cost of tree completion and tree balancing are illustrated for both CPU version and GPU version of the code. When $D_{max}=9$, the final element count is around 0.4 million. Overall, the computational cost of tree completion is trivial compared to that of 2:1 balancing for both CPU version and GPU version of the algorithms. Compared to parallel CPU version of the code (the CPU has 6 physical cores with hyper-threading enabled), we can get a speedup of 5.11 for 2:1 balancing using the newly developed GPU code. In average we allow each thread to create 16 new octants and the rest of the shared memory was used for sorting and removing duplicates within the block of threads. When $D_{max}=10$, the newly generated octants in tree balancing Algorithm~\ref{alg:balance} would exceed the allocated cap.  To resolve this issues, we performed the tree balance algorithm again and a speedup of 3.72 was achieved while the final element count is around 0.9 million. 
Even though, we have to repeat Algorithm~\ref{alg:balance}, the speedup of the GPU code compared to the CPU version is still significant. 
Note that as long as we start CFD simulations with a complete and balanced tree, out of shared memory for refinement would not happen because we only allow any octant to be refined once in per adaptation. 

The speedup of GPU-based tree construction and balancing through this test is very promising.
Admittedly, until this point, we are still steps away from automated initial mesh generation for CFD simulations. One would need to refine the octants near the surface. Another data structure, namely, the GPU-based linear bounded volume hierarchy~\cite{karras2012maximizing}, can be used to accelerate the ray casting algorithms for this purpose. We will further discuss its application to an automated CFD solver in the future. In next section, we are going to focus on the accuracy and efficiency of $p$-adaptive FR using linear trees.
\subsection{Performance study of the adaptive solver}
As a proof of concept, we opt to use the canonical inviscid isentropic vortex propagation problem to study the performance of the adaptive solver with linear trees. The mathematical description of the vortex is formulated as  
\begin{equation}
\begin{cases}
\delta u = -\frac{\alpha}{2\pi}(y-y_0) e ^ {\phi (1-r^2)},\\
\delta v = \frac{\alpha}{2\pi}(x-x_0) e^{\phi (1-r^2)},\\
\delta w = 0,\\
\delta T = -\frac{\alpha^2(\gamma -1)}{16\phi\gamma\pi^2}e^{2\phi(1-r^2)},
\end{cases}
\end{equation}
where $\phi = \frac{1}{2}$, $\alpha = 5$, and $r = (x-x_c(t))^2 + (y-y_c(t))^2$. $(x_c(t),y_c(t))^\top$ is the axis of the vortex center. Initially, the vortex is centered at axis $(0,0)^\top$. The free stream is defined as $(\rho, u, v, w, \mathrm{Ma}) = (1,1,1,0,0.5)$.  A fixed time step size $\Delta t = 0.001$ is used for all simulations.  The minimum element sizes in the mesh $\Delta x_{min}$ were chosen as $25/64$, $25/128$, $25/256$ for all simulations. 

The first set of simulations were run with a 2D computational domain of size $[-12.5,12.5]^2$.  Grid refinement study was performed on both uniform and non-uniform meshes. For AMR with octree, the minimum tree depth was set as  $ D_{min} = 3$ for all 2D simulations and the maximum tree depth would vary from 6,7,8 for different $\Delta x_{min}$.
We used a solution-based adaptation criterion to focus on the computational cost of tree manipulations, data transfer, and solver time. 
The adaptation criterion was defined as the following list.
\begin{itemize}
	\item For an element, if there are more solution points within $\sqrt{(x-x_c(t))^2 + (y-y_c(t))^2} < 5$  than those are $\sqrt{(x-x_c(t))^2 + (y-y_c(t))^2} > 6$, then this element is marked for refinement.
	\item For an element, if there are fewer solution points within $\sqrt{(x-x_c(t))^2 + (y-y_c(t))^2} < 5$   than those are $\sqrt{(x-x_c(t))^2 + (y-y_c(t))^2} > 6$, then this element is marked for coarsening.
	\item Otherwise, the element remains the same.
\end{itemize} 
Adaptation was performed every 10 time steps.~Herein, $L_2$ norm of the density error, which  is defined as
\begin{equation}
L_2(e_\rho) = \sqrt{\int(\rho_{exact} - \rho_{num})^2 dV} ,
\end{equation}
is used for the grid refinement study
and the order of accuracy is defined as 
\begin{equation}
	\text{Order} = log(e_{\rho,1}/e_{\rho,2})/log(\Delta x_{min,1}/\Delta x_{min,2})
\end{equation}
in this work for any Mesh 1 and Mesh 2.

\begin{figure}
 	\centering
	\begin{subfigure}[b]{0.49\textwidth}
	\includegraphics[width=\textwidth]{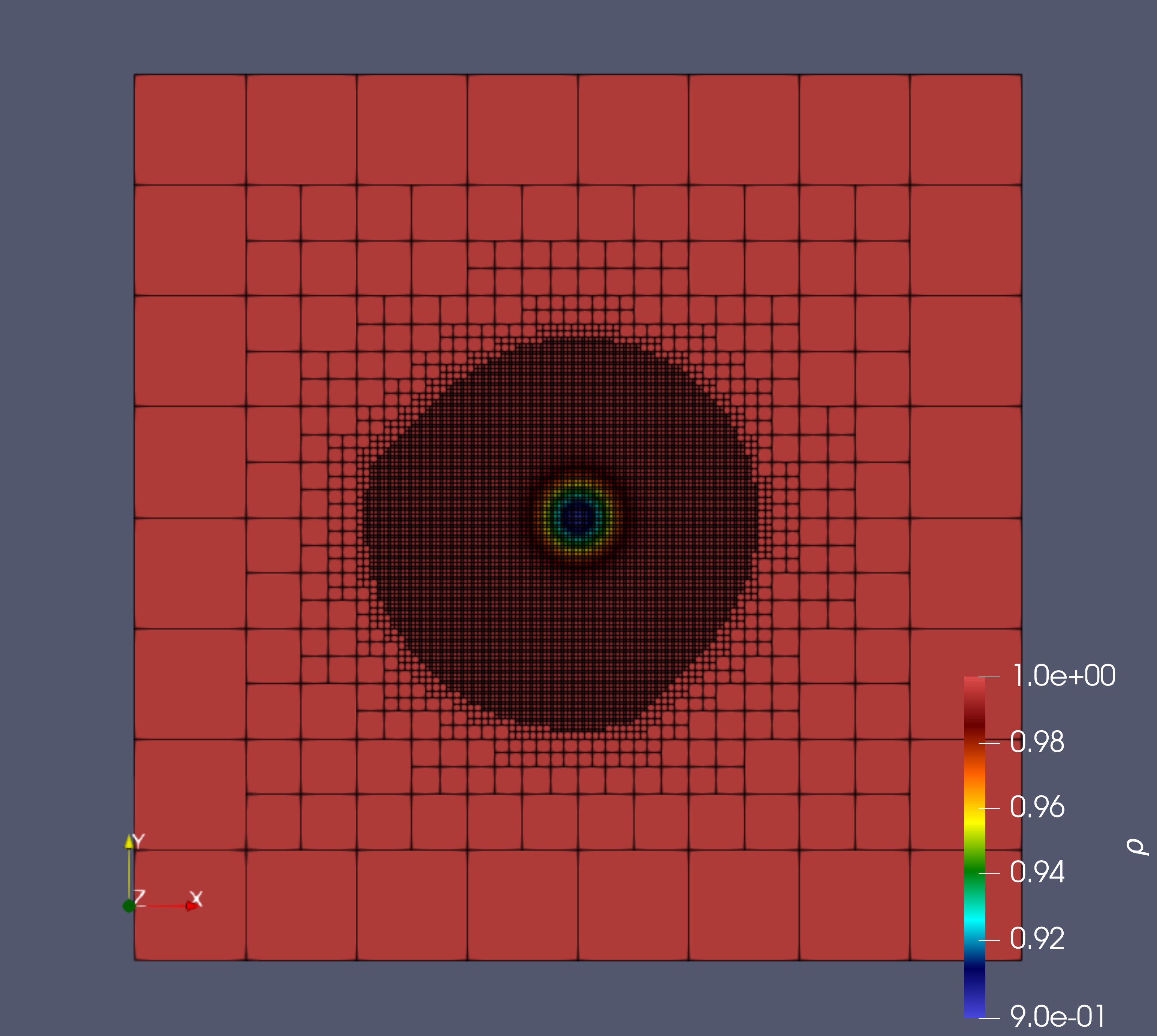}
	\end{subfigure}
	\begin{subfigure}[b]{0.49\textwidth}
	\includegraphics[width=\textwidth]{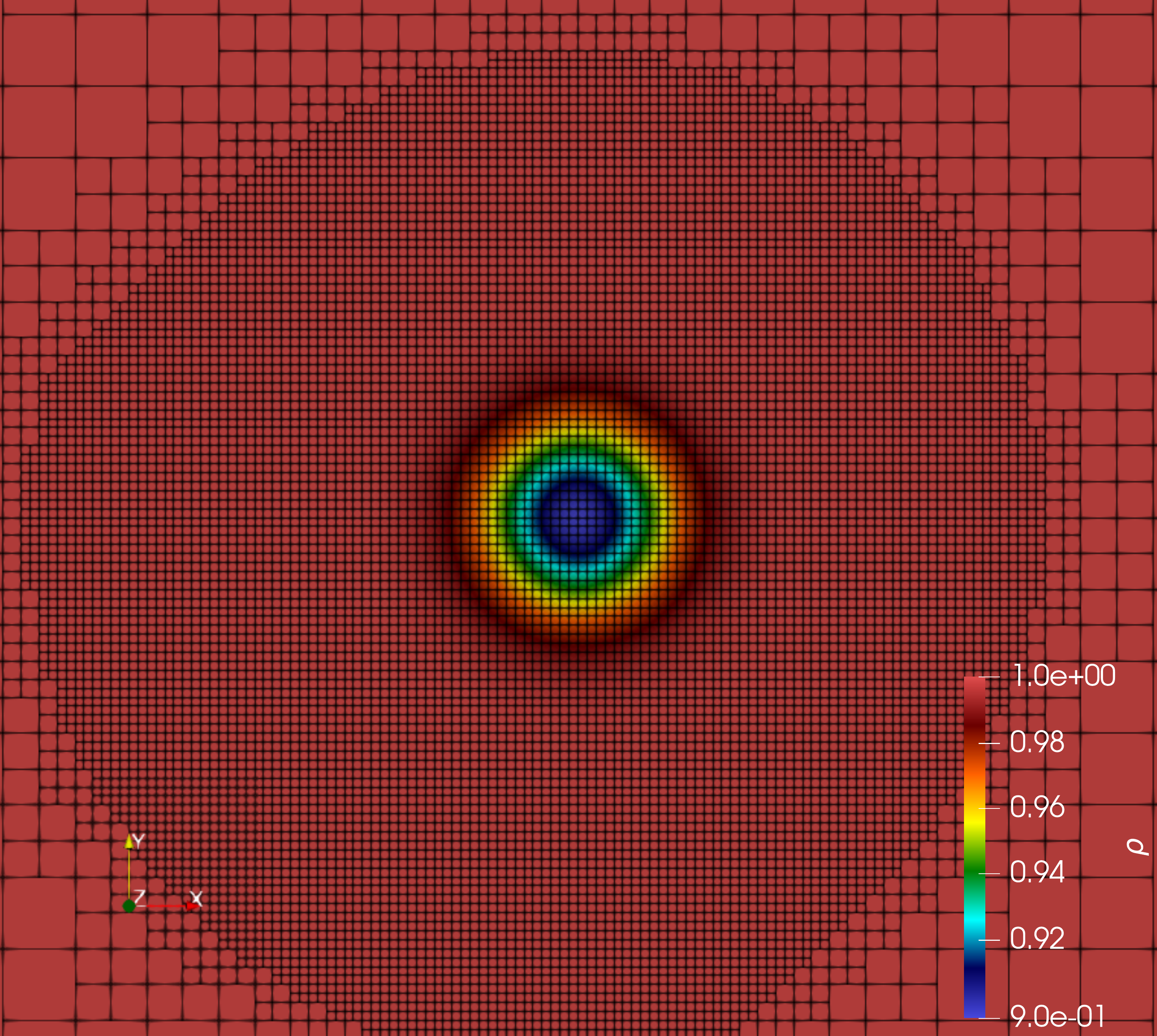}
	\end{subfigure}
	\caption{Final status of the vortex on 2D mesh at $t=25$ for small domain size. $D_{min} =3$ and $D_{max}=8$.}
	\label{fig:vp_p3_2d25}
\end{figure}

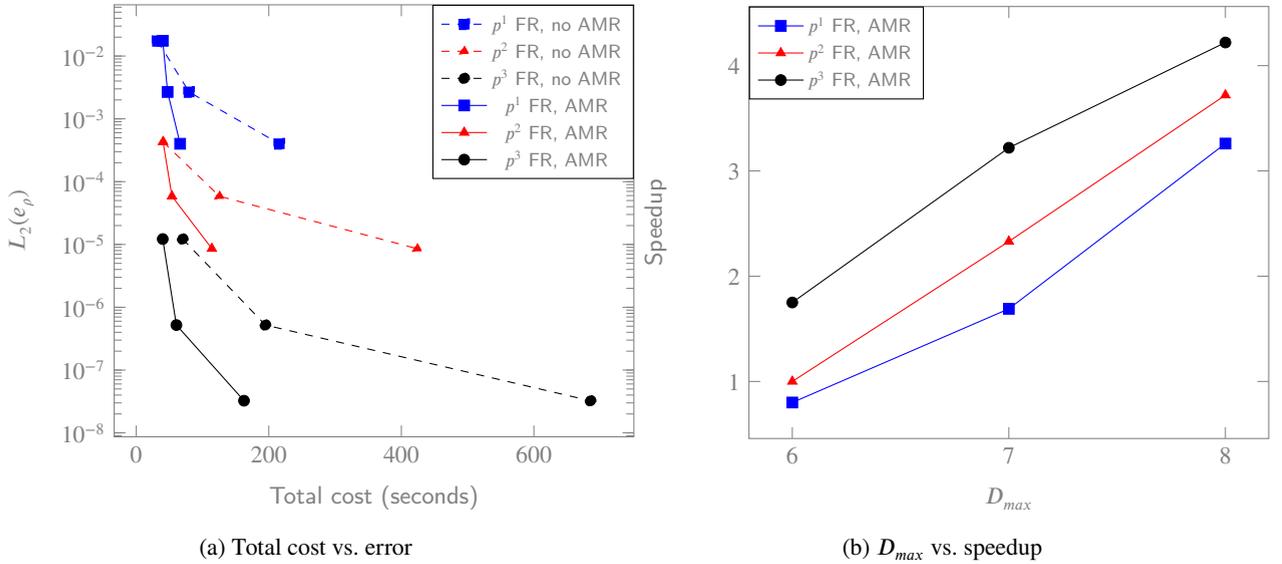
\begin{figure}
	\centering
	\begin{subfigure}[b]{0.49\textwidth}
	\begin{tikzpicture}
		\begin{axis}[
			ymode=log,
			xlabel = Total cost (seconds),
			ylabel = $L_2(e_\rho)$,
			legend style={at={(1,1)},anchor=north east,nodes={scale=0.8, transform shape}}
			]
			\addplot[blue, dashed, mark=square*] table {
				32.02   1.75e-2
				79.66   2.68e-3
				215.87  4.01e-4
			};
			\addlegendentry{$p^1$ FR, no AMR}
			
			\addplot[red, dashed, mark=triangle*] table {
				40.45   4.30e-4
				125.45  5.86e-5
				423.95  8.56e-6
			};
			\addlegendentry{$p^2$ FR, no AMR}
			
			\addplot[black, dashed, mark=otimes*] table {
				70.03   1.22e-5
				194.93  5.21e-7
				685.51  3.24e-8
			};
			\addlegendentry{$p^3$ FR, no AMR}
			
			\addplot[blue,mark=square*] table {
				40.00   1.75e-2
				47.24   2.68e-3
				66.18   4.01e-4
			};
			\addlegendentry{$p^1$ FR, AMR}
			
			\addplot [red, mark=triangle*] table {
				40.45   4.30e-4
				53.82   5.86e-5
				114.05  8.59e-6
			};
			\addlegendentry{$p^2$ FR, AMR}
			
			\addplot [black, mark=otimes*]table {
				39.94  1.22e-5
				60.54  5.21e-7
				162.58 3.25e-8
			};
			\addlegendentry{$p^3$ FR, AMR}
		\end{axis}
	\end{tikzpicture}
	\caption{Total cost vs. error}
	\label{subfig:vp25_error_vs_time}
	\end{subfigure}
		\hfill
		\begin{subfigure}[b]{0.49\textwidth}
		
		\begin{tikzpicture}
			\begin{axis}[
				xlabel = $D_{max}$,
				ylabel = Speedup,
				xtick = {6,7,8},
				legend style={at={(0,1)},anchor=north west,nodes={scale=0.8, transform shape}}
				]
				\addplot[blue, mark=square*] table {
					6   0.80
					7   1.69
					8   3.26
				};
				\addlegendentry{$p^1$ FR, AMR}
				
				\addplot[red,  mark=triangle*] table {
					6   1.00
					7  2.33
					8  3.72
				};
				\addlegendentry{$p^2$ FR, AMR}
				
				\addplot[black, mark=otimes*] table {
					6   1.75
					7  3.22
					8  4.22
				};
				\addlegendentry{$p^3$ FR, AMR}
				
			\end{axis}
		\end{tikzpicture}
	\caption{$D_{max}$ vs. speedup}
	\label{subfig:vp25_speedup}
	\end{subfigure}
\caption{Runtime vs. error for FR with/without AMR for simulations run on 2D domain $[-12.5,12.5]^2$.}
\label{fig:cost_width25}
\end{figure}

\begin{figure}
	\centering
	\begin{subfigure}[b]{0.45\textwidth}
	\begin{tikzpicture}	
		\begin{axis} [ybar = .05cm,
			bar width = 5pt,
			ymin=0,
			ymax=170,
			xmin=5,
			xmax=9,
			xlabel = $D_{max}$,
			ylabel = Runtime (seconds),
			xtick = {6,7,8},
			legend style={at={(0,1)},anchor=north west,nodes={scale=0.8, transform shape}}
			]
			
			\addplot [draw = black,
			line width = .4mm,
			pattern = {north west lines},
			pattern color = black
			] coordinates {(6, 39.9954) (7, 47.2435) (8, 66.181) };
			
			\addplot [draw = blue,
			semithick,
			pattern = {horizontal lines},
			pattern color = blue
			]   coordinates {(6,29.7747) (7,32.2435) (8,48.5595) };
			
			\addplot [draw = red,
			semithick,
			pattern = crosshatch,
			pattern color = red
			]   coordinates {(6,10.2166) (7,14.7417) (8,17.6169) };
			
			\addplot [draw = orange,
			semithick,
			pattern = {north east lines},
			pattern color = orange
			]   coordinates {(6,6.00274) (7,9.8545) (8,12.2898) };
			\addplot [draw = black,
			semithick,
			fill = black
			]   coordinates {(6,1.88514) (7,2.57274) (8,2.9787) };
			\addplot [draw = blue,
			semithick,
			fill = blue
			]   coordinates {(6,2.32873) (7,2.3145) (8,2.34837) };
			\legend {Total, Solver, Adaptation, Tree manipulations, Data transfer, Face connectivity};
		\end{axis}
	\end{tikzpicture}
	\caption{$p^1$ FR with AMR}
	\label{subfig:vp25_comp_p1}
	\end{subfigure}
	\hfill
	\begin{subfigure}[b]{0.45\textwidth}
	\begin{tikzpicture}	
		\begin{axis} [ybar = .05cm,
			bar width = 5pt,
			ymin=0,
			ymax=170,
			xmin=5,
			xmax=9,
			xlabel = $D_{max}$,
			ylabel = Runtime (seconds),
			xtick = {6,7,8},
			legend style={at={(0,1)},anchor=north west,nodes={scale=0.8, transform shape}}
			]
			
			\addplot [draw = black,
			line width = .4mm,
			pattern = {north west lines},
			pattern color = black
			] coordinates {(6, 40.4521) (7, 53.818) (8, 114.05) };
			
			\addplot [draw = blue,
			semithick,
			pattern = {horizontal lines},
			pattern color = blue
			]   coordinates {(6,30.5383) (7,39.0316) (8,91.1745) };
			
			\addplot [draw = red,
			semithick,
			pattern = crosshatch,
			pattern color = red
			]   coordinates {(6,9.90976) (7,14.7821) (8,22.8697) };
			
			\addplot [draw = orange,
			semithick,
			pattern = {north east lines},
			pattern color = orange
			]   coordinates {(6,5.8215) (7,9.76429) (8,15.4939) };
			\addplot [draw = black,
			semithick,
			fill = black
			]   coordinates {(6,1.87055) (7,2.70364) (8,4.3439) };
			\addplot [draw = blue,
			semithick,
			fill = blue
			]   coordinates {(6,2.21772) (7,2.31415) (8,3.03188) };
		\end{axis}
	\end{tikzpicture}
	\caption{$p^2$ FR with AMR}
    \label{subfig:vp25_comp_p2}
	\end{subfigure}
	\begin{subfigure}[b]{0.45\textwidth}
	\begin{tikzpicture}	
		\begin{axis} [ybar = .05cm,
			bar width = 5pt,
			ymin=0,
			ymax=170,
			xmin=5,
			xmax=9,
			xlabel = $D_{max}$,
			ylabel = Runtime (seconds),
			xtick = {6,7,8},
			legend style={at={(0,1)},anchor=north west,nodes={scale=0.8, transform shape}}
			]
			
			\addplot [draw = black,
			line width = .4mm,
			pattern = {north west lines},
			pattern color = black
			] coordinates {(6, 39.9394) (7, 60.539) (8, 162.58) };
			
			\addplot [draw = blue,
			semithick,
			pattern = {horizontal lines},
			pattern color = blue
			]   coordinates {(6,30.4705) (7,45.9371) (8,137.35) };
			
			\addplot [draw = red,
			semithick,
			pattern = crosshatch,
			pattern color = red
			]   coordinates {(6,9.46483) (7,14.5974) (8,25.2243) };
			
			\addplot [draw = orange,
			semithick,
			pattern = {north east lines},
			pattern color = orange
			]   coordinates {(6,5.53636) (7,9.58954) (8,16.9315) };
			\addplot [draw = black,
			semithick,
			fill = black
			]   coordinates {(6,1.8079) (7,2.75756) (8,5.05476) };
			\addplot [draw = blue,
			semithick,
			fill = blue
			]   coordinates {(6,2.12057) (7,2.25031) (8,3.23799) };
		\end{axis}
	\end{tikzpicture}
	\caption{$p^3$ FR with AMR}
	\label{subfig:vp25_comp_p3}
	\end{subfigure}
    \hfill
	\begin{subfigure}[b]{0.45\textwidth}
	\begin{tikzpicture}	
		\begin{axis}[
			xlabel = $D_{max}$,
			ylabel = $\frac{\text{Adaptation}}{\text{Total cost}}$ (\%),
			xtick = {6,7,8},
			legend style={at={(0,1)},anchor=north west,nodes={scale=0.8, transform shape}}
			]
			\addplot[blue, mark=square*] table {
				6   25.55
				7   31.20
				8   26.62
			};
			\addlegendentry{$p^1$ FR, AMR}
			
			\addplot[red,  mark=triangle*] table {
				6   24.50
				7  27.46
				8  20.01
			};
			\addlegendentry{$p^2$ FR, AMR}
			
			\addplot[black, mark=otimes*] table {
				6   23.69
				7  24.12
				8  15.51
			};
			\addlegendentry{$p^3$ FR, AMR}
			
		\end{axis}
	\end{tikzpicture}
	\caption{$D_{max}$ vs. $\frac{\text{Adaptation}}{\text{Total cost}}$ }
	\label{subfig:vp25_speed_ecnt}
	\end{subfigure}
	
	\caption{Computation cost of different components of the adaptive $p^1$, $p^2$, and $p^3$ on smaller domain.}
\end{figure}
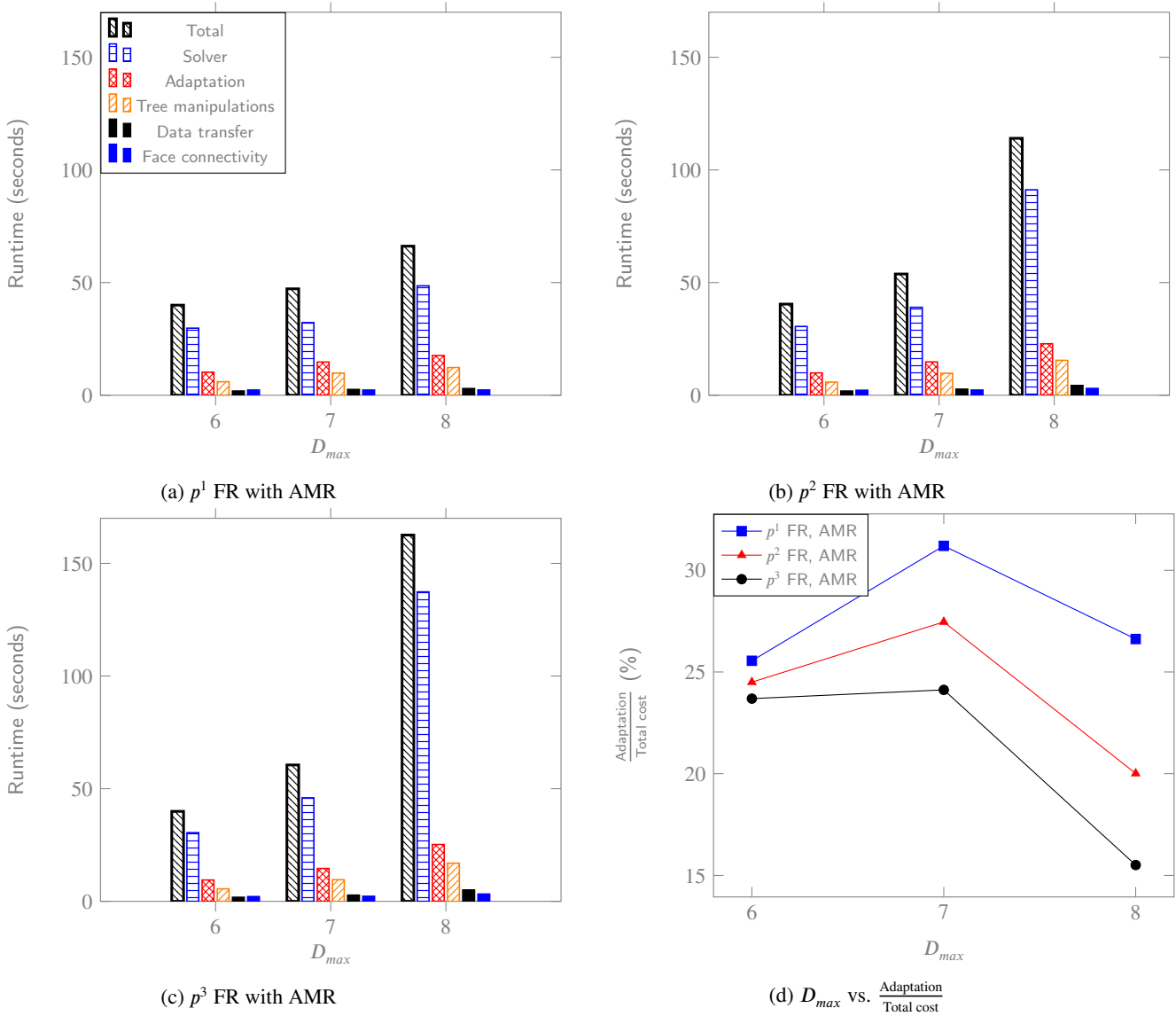

 The statistics of simulations run on uniform meshes are presented in Table~\ref{tab:uni_w25}. Excellent convergence were observed for all polynomial degrees that were tested. 
In Figure~\ref{fig:vp_p3_2d25}, we showcase the final nonuniform mesh of the simulation run using $p^3$ FR. The naive refinement methodology ended up with a very conservative/big refined region.
Statistics of simulations performed on non-uniform meshes can be found in Table~\ref{tab:nonuni_w25}. The adaptation cost in this table includes all aspects related to adaptation, such as  tree manipulations, data transfer and face query. And we only observed trivial $L_2(e_\rho)$ differences between simulations running on uniform and non-uniform meshes.  As expected, the computational cost can be significantly reduced through mesh adaptation. In Figure~\ref{subfig:vp25_error_vs_time}, we present the plots of total cost vs. $L_2(e_\rho)$ for these simulations. As shown in Figure~\ref{subfig:vp25_speedup}, when the polynomial degree increases, the achieved speedup increases as well. And a speedup of 4.22 was achieved for $p^3$ FR when $D_{max}=8$. This is a good testimony of the advantage of using high-order methods with octree-based adaptation techniques and is consistency with what we observed in our previous work of polynomial adaptation~\cite{wang2020dynamically}. For the same polynomial degree, when the mesh was refined, speedup became more significant simply due to the fact the reduction in total element count became more profound. Note that for small problems, the overhead of adaptation was overwhelming such that no or trivial speedup was achieved.
In Figure~\ref{subfig:vp25_comp_p1}, \ref{subfig:vp25_comp_p2}, and \ref{subfig:vp25_comp_p3}, we illustrate runtime of different parts in the adaptive solver including  total cost, total solver cost, and total adaptation cost, where three different components of adaptation cost, namely, tree manipulations, data transfer and face connectivity query are also shown. Note that tree manipulations includes tree adaptation, 2:1 balancing, and finding the mapping between old and new trees. For 2D problems, tree manipulations took the most part of the adaptation time. As the problem size increases, the cost of data transfer, compared to that of the building the face connectivity gradually increases. In Figure~\ref{subfig:vp25_speed_ecnt}, the percentiles of the adaptation cost in the total cost are illustrated. Overall, as the polynomial degree increases, the ratio of $\frac{\text{Adaptation}}{\text{Total cost}}$ will decrease. For $p^3$ polynomials on the finest mesh, adaptation only took around 15\% of the total cost. For $p^1$ polynomials, it could be as much as 31.2\% of the total cost. We argue that performing adaptation every 10  time steps is overkill for explicit time integrators since within such time period the fluid structures would not propagate a significant distance. If the frequency of decreased, the overall overhead of adaptation could be significantly decreased.

\begin{table}[]
	\centering
	\caption{Grid refinement study on domain $[-12.5,12.5]^2$ with 2D uniform meshes.}
	\label{tab:uni_w25}
	\begin{tabular}{lllll}
		\hline
		\multicolumn{5}{l}{Non-adaptive $p^1$ FR}\\
		\hline
		$\Delta x_{min}$&  $L_2(e_\rho)$ & Order  & Total cost (s)  & Element count \\
		$25/64$   & $1.750e-02$    &        & 32.02  & 4096 \\
		$25/128$  & $2.68e-03$     & 2.71   & 79.66  & 16384\\
		$25/256$  & $4.01e-04$     & 2.74   & 215.87 & 65536\\
		\hline
		\multicolumn{5}{l}{Non-adaptive $p^2$ FR}\\
		\hline
		$\Delta x_{min}$&  $L_2(e_\rho)$ & Order  & Total cost (s)  & Element count \\
		$25/64$   & $4.30e-04$    &        & 40.45  & 4096 \\
		$25/128$  & $5.86e-05$    & 2.87   & 125.45 & 16384\\
		$25/256$  & $8.59e-06$    & 2.77   & 423.95 & 65536\\
		\hline
		\multicolumn{5}{l}{Non-adaptive $p^3$ FR}\\
		\hline
		$\Delta x_{min}$&  $L_2(e_\rho)$ & Order  & Total cost (s)  & Element count \\
		$25/64$   & $1.22e-05$    &        & 70.03  & 4096 \\
		$25/128$  & $5.21e-07$    & 4.55   & 194.93 & 16384\\
		$25/256$  & $3.24e-08$    & 4.01   & 685.51 & 65536\\
		\hline
	\end{tabular}
\end{table}

\begin{table}[]
	\centering
	\caption{Grid refinement study on domain $[-12.5,12.5]^2$ with 2D nonuniform meshes.}
	\label{tab:nonuni_w25}
	\begin{tabular}{lllllll}
		\hline
		\multicolumn{7}{l}{Adaptive $p^1$ FR}\\
		\hline
		$\Delta x_{min}$&  $L_2(e_\rho)$ & Order  & Total cost (s) & Adaptation (s) & Element count & $D_{max}$\\
		$25/64$   & $1.75e-02$    &        & 40.00 & 10.22&865   & 6\\
		$25/128$  & $2.68e-03$    & 2.71   & 47.24 & 14.74&2803  & 7\\
		$25/256$  & $4.01e-04$    & 2.74   & 66.18 & 17.62&10519 & 8\\
		\hline
		\multicolumn{7}{l}{Adaptive $p^2$ FR}\\
		\hline
		$\Delta x_{min}$&  $L_2(e_\rho)$ & Order    & Total cost (s) & Adaptation (s) & Element count& $D_{max}$\ \\
		$25/64$   & $4.30e-04$    &       &40.45  & 9.91 &856   & 6\\
		$25/128$  & $5.86e-05$    & 2.87  &53.82  & 14.78&2809  &  7\\
		$25/256$  & $8.59e-06$    & 2.77  &114.05 & 22.87&10513 &8\\
		\hline
		\multicolumn{7}{l}{Adaptive $p^3$ FR}\\
		\hline
		$\Delta x_{min}$&  $L_2(e_\rho)$ & Order  & Total cost (s) &Adaptation (s) & Element count & $D_{max}$\ \\
		$25/64$   & $1.22e-05$    &        & 39.94  & 9.46  &856  & 6\\
		$25/128$  & $5.21e-07$    & 4.55   & 60.54  & 14.60 &2809 &  7\\
		$25/256$  & $3.25e-08$    & 4.00   & 162.58 & 25.22 &10627& 8\\
		\hline
	\end{tabular}
\end{table}

\begin{table}[]
	\centering
	\caption{Grid refinement study on domain $[-50,50]^2$ with 2D uniform meshes.}
	\label{tab:vp100_uni}
	\begin{tabular}{lllll}
		\hline
		\multicolumn{5}{l}{ $p^1$ FR}\\
		\hline
		$\Delta x_{min}$&  $L_2(e_\rho)$ & Order  & Total cost (s)  & Element count \\
		$25/64$    & $4.65e-02$    &        & 870.31    &65536   \\
		$25/128$   & $9.19e-03$    & 2.34   & 3108.43   &262144  \\
		$25/256$*  &               &        & 11142.10  &1048576\\
		\hline
		\multicolumn{5}{l}{ $p^2$ FR}\\
		\hline
		$\Delta x_{min}$&  $L_2(e_\rho)$ & Order  & Total cost (s)  & Element count \\
		$25/64$    & $1.17e-03$    &        & 1747.67   &65536   \\
		$25/128$   & $9.15e-05$    & 3.68   & 6299.62   &262144  \\
		$25/256$*  &               &        & 23417.00  &1048576\\
		\hline
		\multicolumn{5}{l}{ $p^3$ FR}\\
		\hline
		$\Delta x_{min}$&  $L_2(e_\rho)$ & Order  & Total cost (s)  & Element count \\
		$25/64$    & $2.68e-05$    &        & 2720.30 &65536   \\
		$25/128$   & $5.34e-07$    & 5.64   & 9993.02 &262144  \\ 
		\hline
	\end{tabular}\\
\end{table}

\begin{table}[]
	\centering
	\caption{Grid refinement study on domain $[-50,50]^2$ with 2D nonuniform meshes.}
	\label{tab:vp100_nonuni}
	\begin{tabular}{lllllll}
		\hline
		\multicolumn{7}{l}{Adaptive $p^1$ FR}\\
		\hline
		$\Delta x_{min}$&  $L_2(e_\rho)$ & Order  & Total cost (s) & Adaptation (s) & Element count &$D_{max}$\\
		$25/64$   & $4.64e-02$    &        & 182.39 & 56.75 &961   & 8\\
		$25/128$  & $9.19e-03$    & 2.33   & 211.71 & 77.14 &2899  & 9\\
		$25/256$  & $1.28e-03$    & 2.84   & 294.26 & 92.44 &10615 &10\\
		\hline		
		\multicolumn{7}{l}{Adaptive $p^2$ FR}\\
		\hline
		$\Delta x_{min}$&  $L_2(e_\rho)$ & Order  & Total cost (s) & Adaptation (s)  & Element count &$D_{max}$ \\
		$25/64$   & $1.18e-03$  &        & 168.80 & 49.13 &952  & 8\\
		$25/128$  & $9.19e-05$  & 3.68   & 230.35 & 76.90 &2905 & 9\\
		$25/256$  & $9.54e-06$  & 3.27   & 488.48 &115.14 &10609&10\\
		\hline
		\multicolumn{7}{l}{Adaptive $p^3$ FR}\\
		\hline
		$\Delta x_{min}$&  $L_2(e_\rho)$ & Order  & Total cost (s) & Adaptation (s)  & Element count & $D_{max}$ \\
		$25/64$   & $2.68e-05$    &        & 176.38 & 49.69 &952  & 8\\
		$25/128$  & $5.34e-07$    & 5.64   & 273.20 & 75.77&2905 & 9\\
		$25/256$  & $3.46e-08$    & 3.95   & 710.11 & 135.08&10723 &10\\
		\hline
	\end{tabular}
\end{table}

\begin{figure}
 	\centering
	\begin{subfigure}[b]{0.49\textwidth}
	\includegraphics[width=\textwidth]{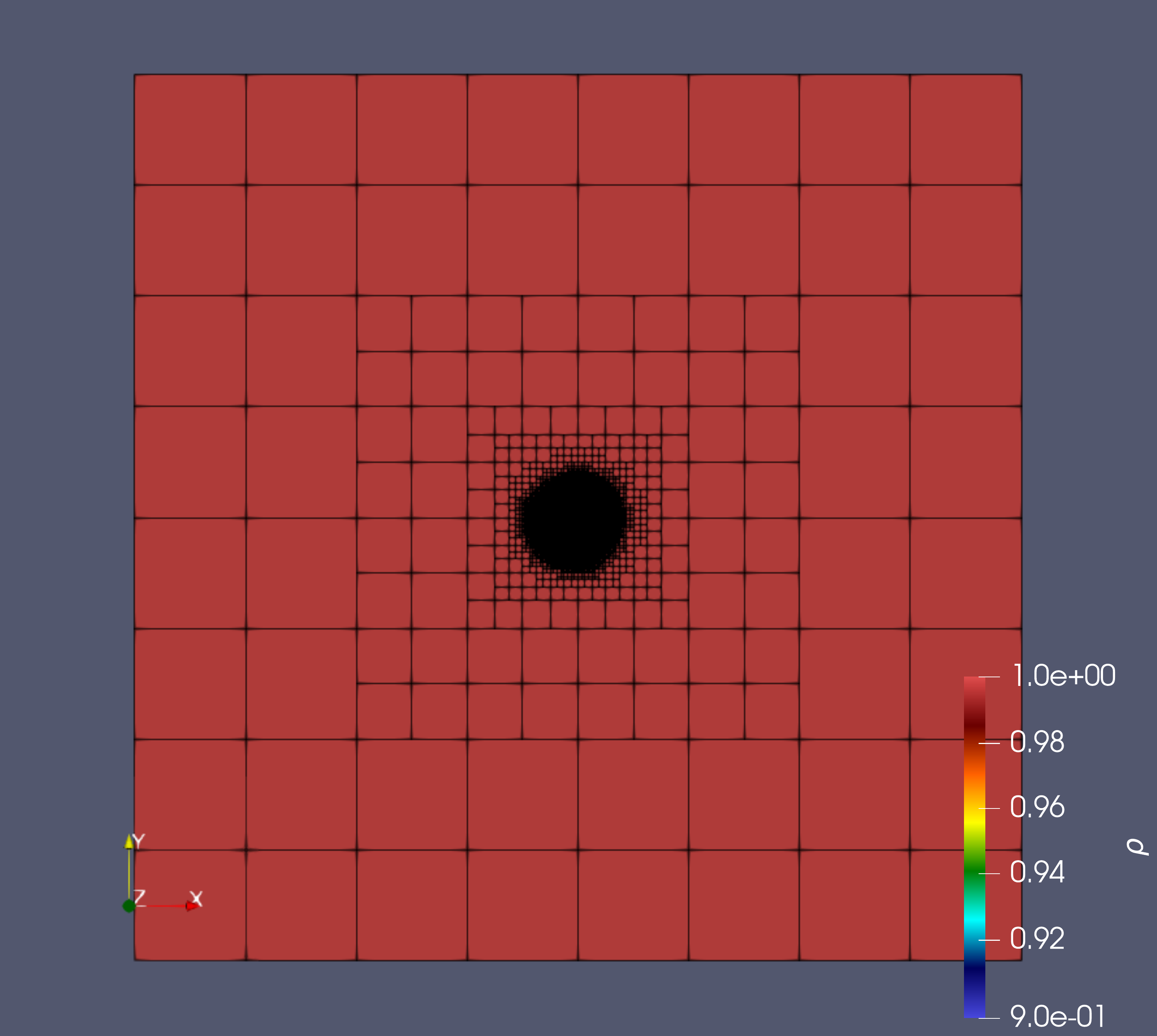}
	\end{subfigure}
	\begin{subfigure}[b]{0.49\textwidth}
	\includegraphics[width=\textwidth]{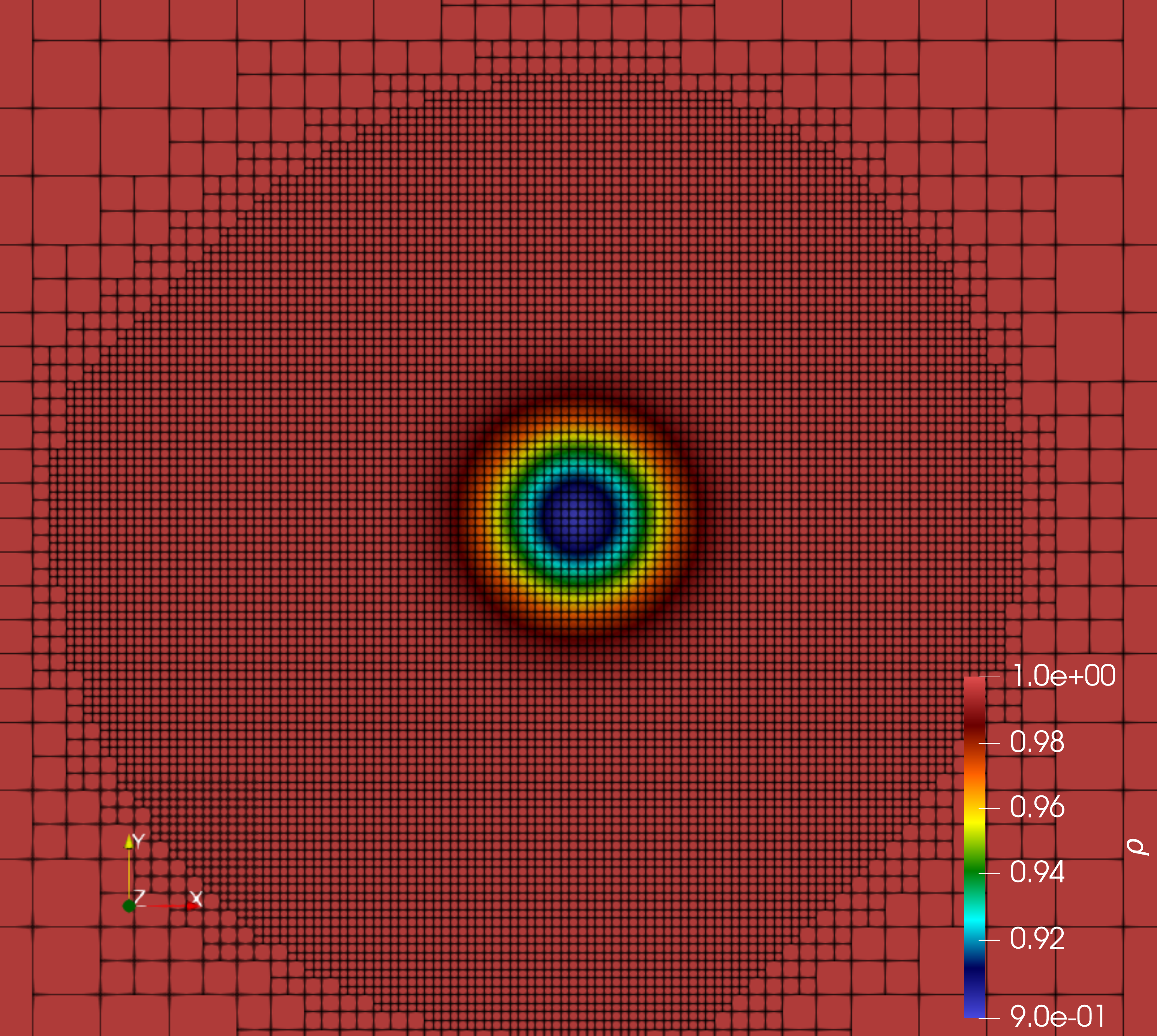}
	\end{subfigure}
	\caption{Final status of the vortex on 2D mesh at $t=100$ for larger domain size using $p^3$ FR. $D_{min} =3$ and $D_{max}=10$.}
	\label{fig:vp_p3_2d100}
\end{figure}

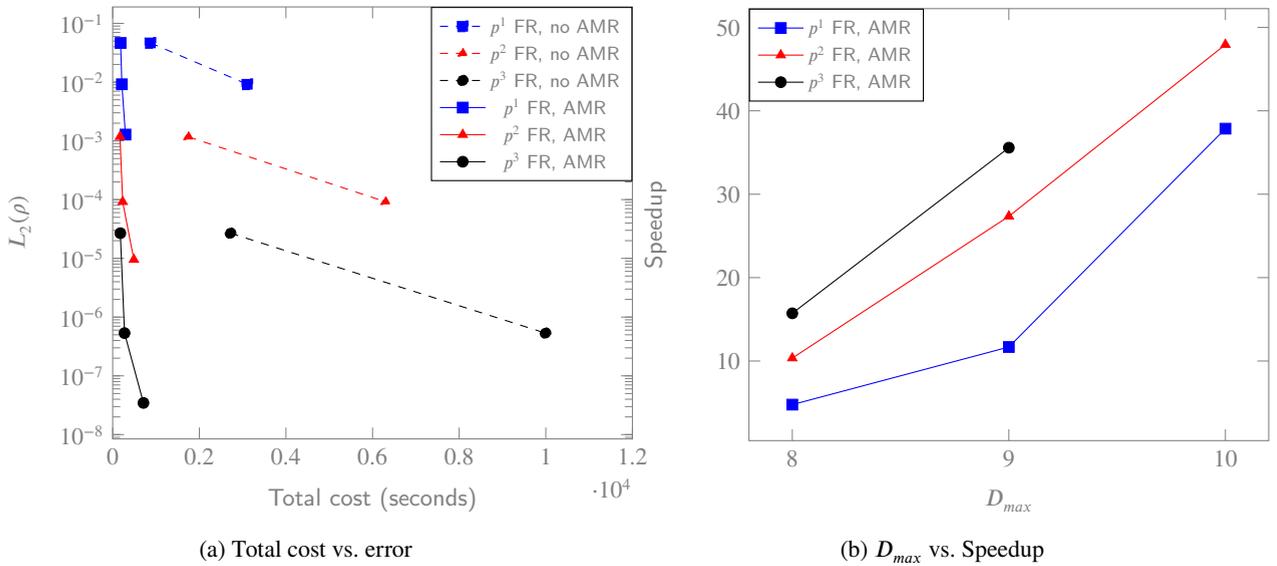
\begin{figure}
	\centering
	\begin{subfigure}[b]{0.49\textwidth}
		\label{subfig:vp100_error_vs_time}
		\begin{tikzpicture}
			\begin{axis}[
				ymode=log,
				xlabel = Total cost (seconds),
				ylabel = $L_2(\rho)$,
				xmin =0,
				xmax = 1.2e+4,
				legend style={at={(1,1)},anchor=north east,nodes={scale=0.8, transform shape}}
				]
				\addplot[blue, dashed, mark=square*] table {
					870.31   4.65e-2
					3108.43  9.19e-3 
				};
				\addlegendentry{$p^1$ FR, no AMR}
				
				\addplot[red, dashed, mark=triangle*] table {
					1747.67   1.17e-3
					6299.62   9.19e-5 
				};
				\addlegendentry{$p^2$ FR, no AMR}
				
				\addplot[black, dashed, mark=otimes*] table {
					2720.30   2.68e-5
					9993.02   5.34e-7 
				};
				\addlegendentry{$p^3$ FR, no AMR}
				
				\addplot[blue,mark=square*] table {
					182.39   4.64e-2
					211.71   9.19e-3
					294.26   1.28e-3
				};
				\addlegendentry{$p^1$ FR, AMR}
				
				\addplot [red, mark=triangle*] table {
					168.80   1.18e-3
					230.35   9.19e-5
					488.48   9.54e-6
				};
				\addlegendentry{$p^2$ FR, AMR}
				
				\addplot [black, mark=otimes*]table {
					176.38  2.68e-5
					273.20  5.34e-7
					710.11  3.46e-8
				};
				\addlegendentry{$p^3$ FR, AMR}
			\end{axis}
		\end{tikzpicture}
	\caption{Total cost vs. error}
	\end{subfigure}
	\hfill
	\begin{subfigure}[b]{0.49\textwidth}
		\label{subfig:vp100_speedup}
		\begin{tikzpicture}
			\begin{axis}[
				ylabel = Speedup,
				xlabel = $D_{max}$,
				xtick = {8,9,10},
				legend style={at={(0,1)},anchor=north west,nodes={scale=0.8, transform shape}}
				]

				\addplot[blue,mark=square*] table {
					8   4.77
					9   11.68
					10   37.86
				};
				\addlegendentry{$p^1$ FR, AMR}
				
				\addplot [red, mark=triangle*] table {
					8  10.35
					9   27.35
					10  47.94
				};
				\addlegendentry{$p^2$ FR, AMR}
				
				\addplot [black, mark=otimes*]table {
					8  15.72
					9  35.58
				};
				\addlegendentry{$p^3$ FR, AMR}
			\end{axis}
		\end{tikzpicture}
        \caption{$D_{max}$ vs. Speedup}
	\end{subfigure}
	\caption{Runtime vs. error for FR with/without AMR for simulations run on 2D domain $[-50,50]^2$.}
        \label{fig:vp100_speedup}
\end{figure}

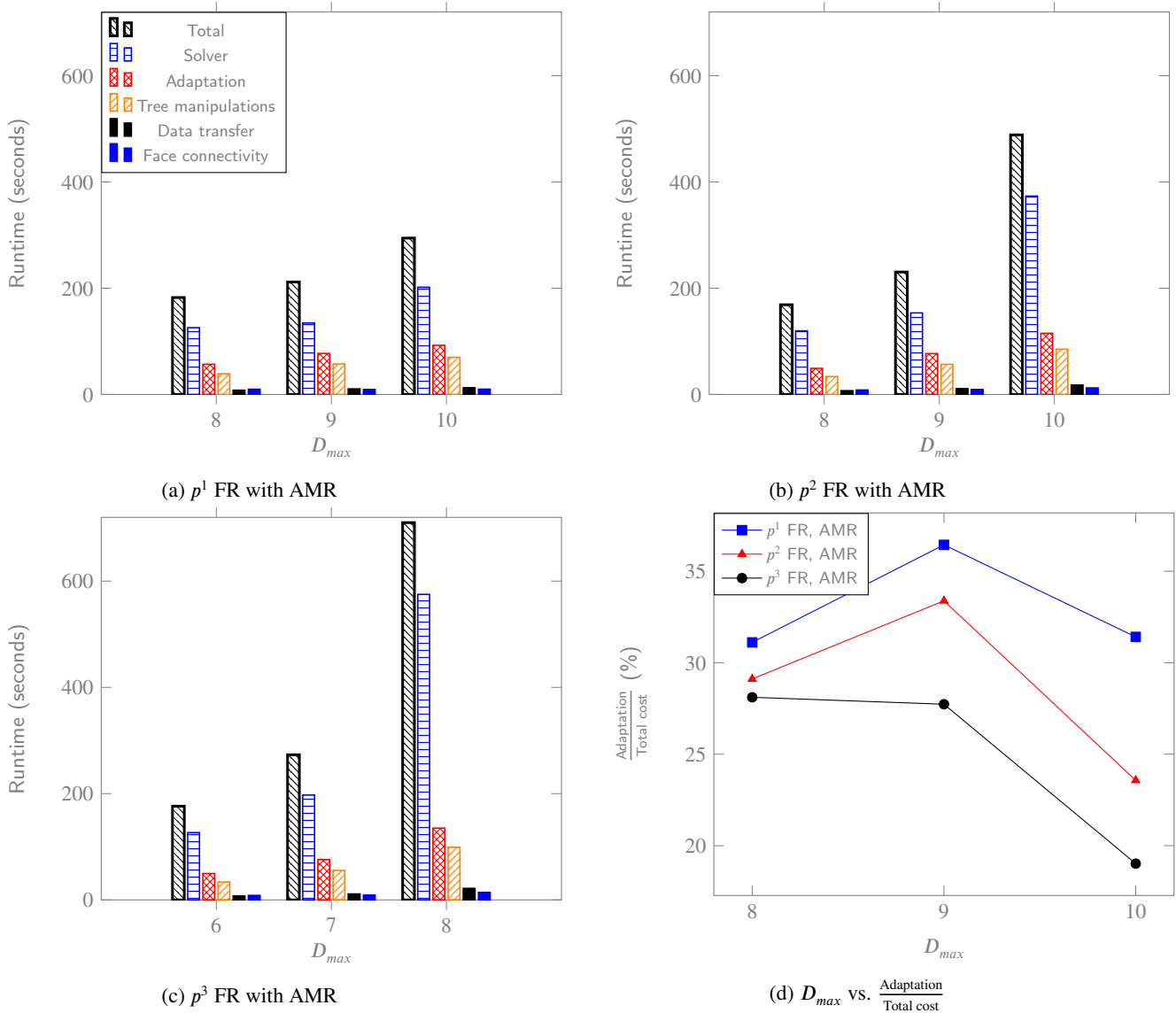
\begin{figure}
	\centering
	\begin{subfigure}[b]{0.45\textwidth}
		\begin{tikzpicture}	
			\begin{axis} [ybar = .05cm,
				bar width = 5pt,
				ymin=0,
				ymax=720,
				xmin=7,
				xmax=11,
				xlabel = $D_{max}$,
				ylabel = Runtime (seconds),
				xtick = {8,9,10},
				legend style={at={(0,1)},anchor=north west,nodes={scale=0.8, transform shape}}
				]
				
				\addplot [draw = black,
				line width = .4mm,
				pattern = {north west lines},
				pattern color = black
				] coordinates {(8, 182.388) (9, 211.714) (10, 294.26) };
				
				\addplot [draw = blue,
				semithick,
				pattern = {horizontal lines},
				pattern color = blue
				]   coordinates {(8,125.618) (9,134.56) (10,201.806) };
				
				\addplot [draw = red,
				semithick,
				pattern = crosshatch,
				pattern color = red
				]   coordinates {(8,56.7525) (9,77.1366) (10,92.4353) };
				
				\addplot [draw = orange,
				semithick,
				pattern = {north east lines},
				pattern color = orange
				]   coordinates {(8,38.9289) (9,57.2476) (10,69.7881) };
				\addplot [draw = black,
				semithick,
				fill = black
				]   coordinates {(8,7.98592) (9,10.4657) (10,12.706) };
				\addplot [draw = blue,
				semithick,
				fill = blue
				]   coordinates {(8,9.83767) (9,9.42325) (10,9.9412) };
				\legend {Total, Solver, Adaptation, Tree manipulations, Data transfer, Face connectivity};
			\end{axis}
		\end{tikzpicture}
		\caption{$p^1$ FR with AMR}
		\label{subfig:vp100_comp_p1}
	\end{subfigure}
	\hfill
	\begin{subfigure}[b]{0.45\textwidth}
		\begin{tikzpicture}	
			\begin{axis} [ybar = .05cm,
				bar width = 5pt,
				ymin=0,
				ymax=720,
				xmin=7,
				xmax=11,
				xlabel = $D_{max}$,
				ylabel = Runtime (seconds),
				xtick = {8,9,10},
				legend style={at={(0,1)},anchor=north west,nodes={scale=0.8, transform shape}}
				]
				
				\addplot [draw = black,
				line width = .4mm,
				pattern = {north west lines},
				pattern color = black
				] coordinates {(8, 168.797) (9, 230.352) (10, 488.482) };
				
				\addplot [draw = blue,
				semithick,
				pattern = {horizontal lines},
				pattern color = blue
				]   coordinates {(8,119.657) (9,153.428) (10,373.321) };
				
				\addplot [draw = red,
				semithick,
				pattern = crosshatch,
				pattern color = red
				]   coordinates {(8,49.125) (9,76.9045) (10,115.139) };
				
				\addplot [draw = orange,
				semithick,
				pattern = {north east lines},
				pattern color = orange
				]   coordinates {(8,33.6758) (9,56.5947) (10,85.1652) };
				\addplot [draw = black,
				semithick,
				fill = black
				]   coordinates {(8,7.07162) (9,10.9418) (10,17.6433) };
				\addplot [draw = blue,
				semithick,
				fill = blue
				]   coordinates {(8,8.37763) (9,9.36799) (10,12.3301) };
			\end{axis}
		\end{tikzpicture}
		\caption{$p^2$ FR with AMR}
		\label{subfig:vp100_comp_p2}
	\end{subfigure}
	\begin{subfigure}[b]{0.45\textwidth}
		\begin{tikzpicture}	
			\begin{axis} [ybar = .05cm,
				bar width = 5pt,
				ymin=0,
				ymax=720,
				xmin=5,
				xmax=9,
				xlabel = $D_{max}$,
				ylabel = Runtime (seconds),
				xtick = {6,7,8},
				legend style={at={(0,1)},anchor=north west,nodes={scale=0.8, transform shape}}
				]
				
				\addplot [draw = black,
				line width = .4mm,
				pattern = {north west lines},
				pattern color = black
				] coordinates {(6, 176.382) (7, 273.201) (8, 710.112) };
				
				\addplot [draw = blue,
				semithick,
				pattern = {horizontal lines},
				pattern color = blue
				]   coordinates {(6,126.68) (7,197.414) (8,575.009) };
				
				\addplot [draw = red,
				semithick,
				pattern = crosshatch,
				pattern color = red
				]   coordinates {(6,49.6856) (7,75.7712) (8,135.078) };
				
				\addplot [draw = orange,
				semithick,
				pattern = {north east lines},
				pattern color = orange
				]   coordinates {(6,33.8092) (7,55.4552) (8,99.0834) };
				\addplot [draw = black,
				semithick,
				fill = black
				]   coordinates {(6,7.40352) (7,11.2142) (8,21.7472) };
				\addplot [draw = blue,
				semithick,
				fill = blue
				]   coordinates {(6,8.47285) (7,9.10176) (8,14.2473) };
			\end{axis}
		\end{tikzpicture}
		\caption{$p^3$ FR with AMR}
		\label{subfig:vp100_comp_p3}
	\end{subfigure}
	\hfill
	\begin{subfigure}[b]{0.45\textwidth}
		\begin{tikzpicture}	
			\begin{axis}[
				xlabel = $D_{max}$,
				ylabel = $\frac{\text{Adaptation}}{\text{Total cost}}$ (\%),
				xtick = {8,9,10},
				legend style={at={(0,1)},anchor=north west,nodes={scale=0.8, transform shape}}
				]
				\addplot[blue, mark=square*] table {
					8   31.11
					9   36.44
					10   31.41
				};
				\addlegendentry{$p^1$ FR, AMR}
				
				\addplot[red,  mark=triangle*] table {
					8   29.11
					9  33.38
					10  23.57
				};
				\addlegendentry{$p^2$ FR, AMR}
				
				\addplot[black, mark=otimes*] table {
					8   28.11
					9  27.73
					10  19.02
				};
				\addlegendentry{$p^3$ FR, AMR}
				
			\end{axis}
		\end{tikzpicture}
		\caption{$D_{max}$ vs. $\frac{\text{Adaptation}}{\text{Total cost}}$ }
		\label{subfig:vp100_speed_ecnt}
	\end{subfigure}
	
	\caption{Computation cost of different components of adaptive $p^1$, $p^2$, and $p^3$ FR on domain $[-50,50]^2$.}
 \label{fig:vp100_components}
\end{figure}

\begin{figure}
		\begin{subfigure}[b]{0.49\textwidth}
		\label{subfig:d678_e_vs_t}
		\begin{tikzpicture}
			\begin{axis}[
				xlabel = Element count,
				ylabel = Adaptation cost,
                ymin = 0,
                ymax = 35,
				legend style={at={(0,1)},anchor=north west,nodes={scale=0.8, transform shape}}
				]
				
				\addplot[blue,mark=square*] table {
					865     10.22 
					2803    14.74
					10519   17.62
				};
				\addlegendentry{$p^1$ FR, AMR}
				
				\addplot [red, mark=triangle*] table {
					856    9.91
					2809   14.78
					10513  22.87
				};
				\addlegendentry{$p^2$ FR, AMR}
				
				\addplot [black, mark=otimes*]table {
					856    9.46
					2809   14.60
					10627  22.52
				};
				\addlegendentry{$p^3$ FR, AMR}
			\end{axis}
		\end{tikzpicture}
	\caption{$D_{max} = 6,7,8$ and $D_{min} = 3$.}
	\end{subfigure}
	\hfill
	\begin{subfigure}[b]{0.49\textwidth}
		\label{subfig:d8910_e_vs_t}
		\begin{tikzpicture}
			\begin{axis}[
				xlabel = Element count,
				ylabel = Adaptation cost,
                ymin = 0,
                ymax = 35,
				legend style={at={(0,1)},anchor=north west,nodes={scale=0.8, transform shape}}
				]
				
				\addplot[blue,mark=square*] table {
					961   14.1875
					2899   19.285
					10615   23.11
				};
				\addlegendentry{$p^1$ FR, AMR}
				
				\addplot [red, mark=triangle*] table {
					952   12.2825
					2905   19.225
					10629  28.785
				};
				\addlegendentry{$p^2$ FR, AMR}
				
				\addplot [black, mark=otimes*]table {
					952  12.4225
					2905  18.9425
					10723 33.77
				};
				\addlegendentry{$p^3$ FR, AMR}
			\end{axis}
		\end{tikzpicture}
	\caption{$D_{max} = 8,9,10$ and $D_{min}=3$. Cost normalized by 4.}
	\end{subfigure}
	\caption{Effect of tree depth on the cost of adaptation.}
	\label{fig:d_effect}
\end{figure}
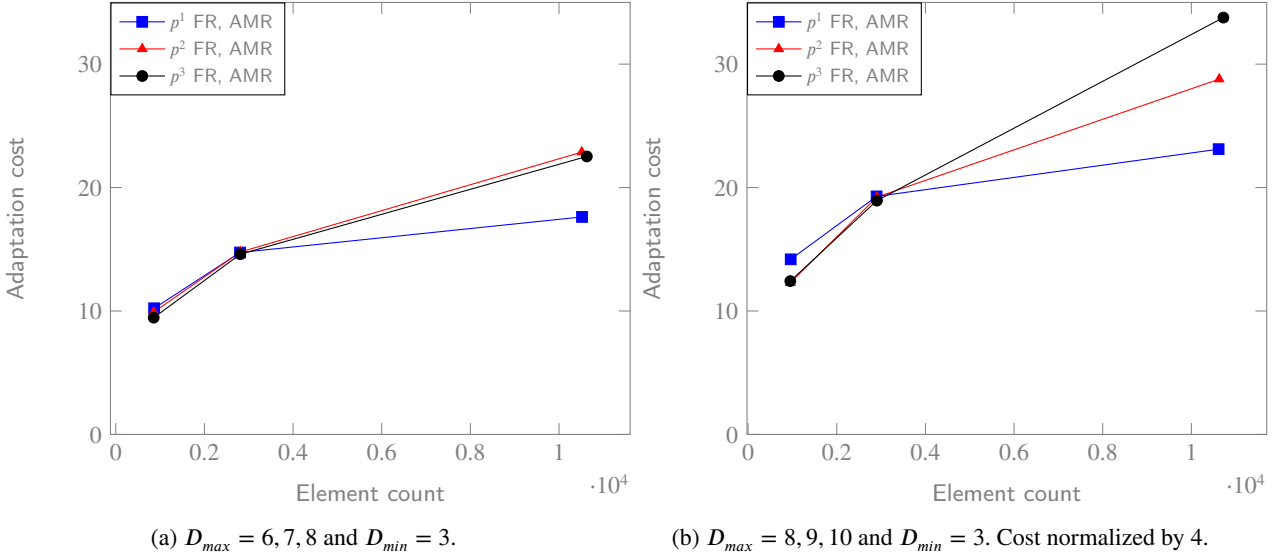

The second test we performed was on a larger computational domain, such that the vortex could propagate a longer distance within one period and larger maximum tree depths could be used to test the efficiency of our adaptive solver. 
The computational domain was within $[-100,100]^2$ and the final status of the mesh and vortex are shown in Figure~\ref{fig:vp_p3_2d100}. Overall statistics of the simulations on uniform meshes and nonuniform meshes are documented in Table~\ref{tab:vp100_uni} and Table~\ref{tab:vp100_nonuni}.
Note that in Table~\ref{tab:vp100_uni} we only run the simulation on the uniform mesh with $\Delta x_{min}=25/256$ for 5 time units and multiplied the cost by 20 to infer to the total cost for $p^1$ and $p^2$ FR. This was due to that the dusted laptop suffered from over-heating when simulation run for hours. We could not run the simulation on the finest uniform mesh with $p^3$ FR because of not enough GPU memory. 
As impressive as the performance of GPU-based solver already is on uniform meshes, the acceleration that AMR offers  for this particular experiment, shown in Figure~\ref{fig:vp100_speedup},  is astonishing with a speedup of 49 for $p^2$ polynomials on the finest mesh. In Figure~\ref{fig:vp100_components}, bar charts of different components of the entire solver are plotted. The trends of different components are similar to what are observed for simulations run on the smaller domain. The overhead of adaptation compared to the total cost slightly increased.  
In Figure~\ref{fig:d_effect}, we present the effect of tree depth on the adaptation cost. When larger tree depth is allowed, in Algorithm~\ref{alg:balance}, one need to perform balancing for more levels. Therefore, it is reasonable to see that the normalized adaptation cost of simulations run using $D_{max}=10$ was larger than that using $D_{max}=8$ given that the element count were close.

\begin{figure}
	\centering
	\includegraphics[width=0.5\textwidth]{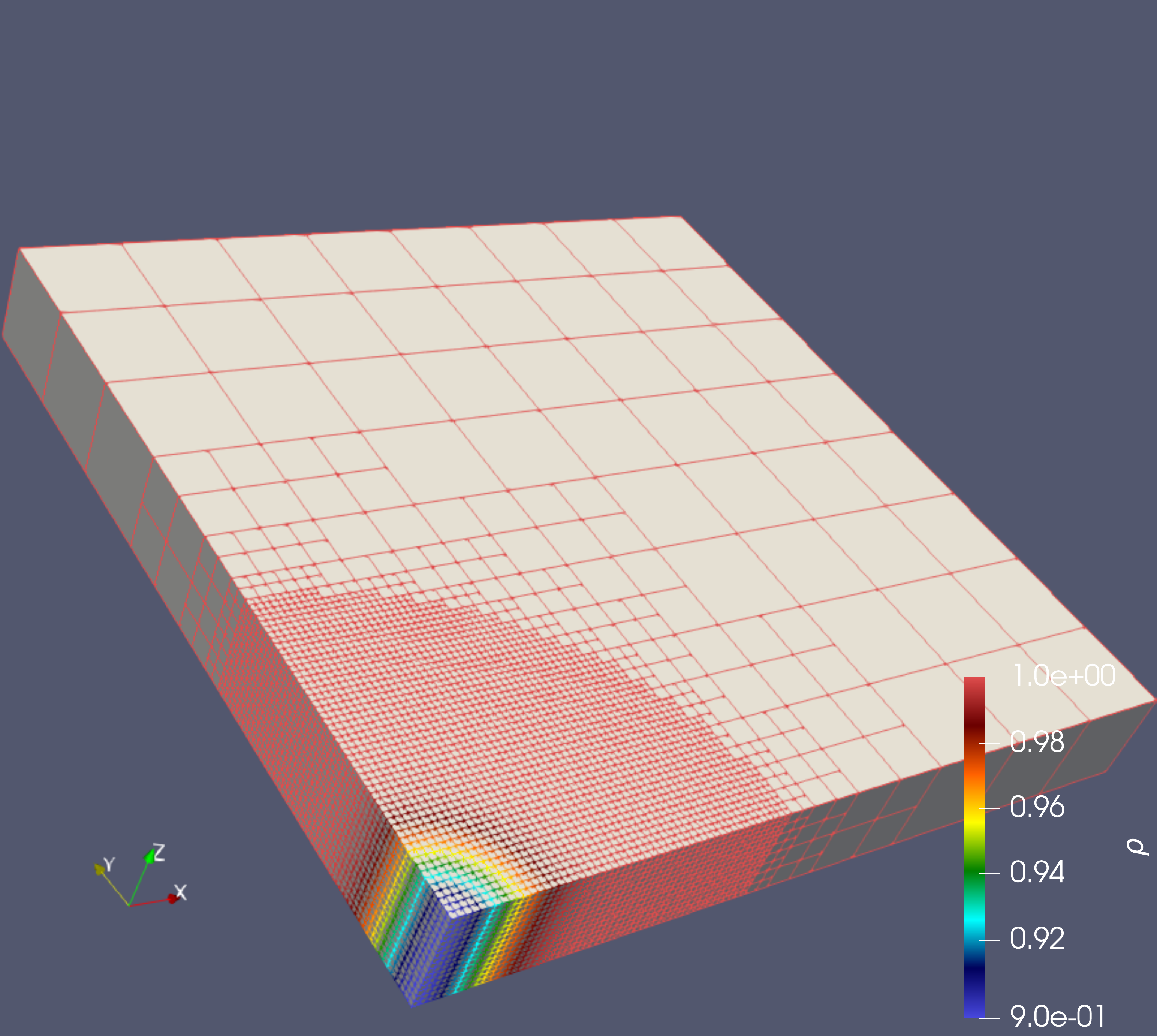}
	\caption{Final status of the top-right quarter of the vortex on the 3D nonuniform mesh at $t=25$ using $p^2$ FR. $D_{min}=4$ and $D_{max}=8$.}
	\label{fig:vp_3d_mesh}
\end{figure}

\begin{figure}
	\centering
	\begin{subfigure}[b]{0.45\textwidth}
		\begin{tikzpicture}	
			\begin{axis} [ybar = .05cm,
				bar width = 5pt,
				ymin=0,
				ymax=4200,
				xmin=5,
				xmax=9,
				xlabel = $D_{max}$,
				ylabel = Runtime (seconds),
				xtick = {6,7,8},
				legend style={at={(0,1)},anchor=north west,nodes={scale=0.8, transform shape}}
				]
				
				\addplot [draw = black,
				line width = .4mm,
				pattern = {north west lines},
				pattern color = black
				] coordinates {(6, 59.0526) (7, 244.392) (8, 1473.02) };
				
				\addplot [draw = blue,
				semithick,
				pattern = {horizontal lines},
				pattern color = blue
				]   coordinates {(6,47.69) (7,217.95) (8,1411.29) };
				
				\addplot [draw = red,
				semithick,
				pattern = crosshatch,
				pattern color = red
				]   coordinates {(6,11.3575) (7,26.4346) (8,61.7185) };
				
				\addplot [draw = orange,
				semithick,
				pattern = {north east lines},
				pattern color = orange
				]   coordinates {(6,6.04531) (7,15.4534) (8,37.9543) };
				\addplot [draw = black,
				semithick,
				fill = black
				]   coordinates {(6,2.83718) (7,6.199) (8,14.9726) };
				\addplot [draw = blue,
				semithick,
				fill = blue
				]   coordinates {(6,2.47499) (9,4.78222) (8,8.79172) };
				\legend {Total, Solver, Adaptation, Tree manipulations, Data transfer, Face connectivity};
			\end{axis}
		\end{tikzpicture}
		\caption{$p^1$ FR with AMR}
		\label{subfig:vp25_3d_comp_p1}
	\end{subfigure}
	\hfill
	\begin{subfigure}[b]{0.45\textwidth}
		\begin{tikzpicture}	
			\begin{axis} [ybar = .05cm,
				bar width = 5pt,
				ymin=0,
				ymax=4200,
				xmin=5,
				xmax=9,
				xlabel = $D_{max}$,
				ylabel = Runtime (seconds),
				xtick = {6,7,8},
				legend style={at={(0,1)},anchor=north west,nodes={scale=0.8, transform shape}}
				]
				
				\addplot [draw = black,
				line width = .4mm,
				pattern = {north west lines},
				pattern color = black
				] coordinates {(6, 122.557) (7, 607.23) (8, 4193.54) };
				
				\addplot [draw = blue,
				semithick,
				pattern = {horizontal lines},
				pattern color = blue
				]   coordinates {(6,108.335) (7,579.995) (8,4112.14) };
				
				\addplot [draw = red,
				semithick,
				pattern = crosshatch,
				pattern color = red
				]   coordinates {(6,14.2161) (7,27.2276) (8,81.3875) };
				
				\addplot [draw = orange,
				semithick,
				pattern = {north east lines},
				pattern color = orange
				]   coordinates {(6,7.5146) (7,15.2127) (8,40.6145) };
				\addplot [draw = black,
				semithick,
				fill = black
				]   coordinates {(6,3.80409) (7,7.90066) (8,31.9854) };
				\addplot [draw = blue,
				semithick,
				fill = blue
				]   coordinates {(6,2.89738) (7,4.1143) (8,8.78749) };
			\end{axis}
		\end{tikzpicture}
		\caption{$p^2$ FR with AMR}
		\label{subfig:vp25_3d_comp_p2}
	\end{subfigure}
	\begin{subfigure}[b]{0.45\textwidth}
		\begin{tikzpicture}	
			\begin{axis} [ybar = .05cm,
				bar width = 5pt,
				ymin=0,
				ymax=4200,
				xmin=5,
				xmax=9,
				xlabel = $D_{max}$,
				ylabel = Runtime (seconds),
				xtick = {6,7,8},
				legend style={at={(0,1)},anchor=north west,nodes={scale=0.8, transform shape}}
				]
				
				\addplot [draw = black,
				line width = .4mm,
				pattern = {north west lines},
				pattern color = black
				] coordinates {(6, 226.831) (7, 1301.09)  };
				
				\addplot [draw = blue,
				semithick,
				pattern = {horizontal lines},
				pattern color = blue
				]   coordinates {(6,209.507) (7,1264.36)  };
				
				\addplot [draw = red,
				semithick,
				pattern = crosshatch,
				pattern color = red
				]   coordinates {(6,17.3179) (7,36.7177)  };
				
				\addplot [draw = orange,
				semithick,
				pattern = {north east lines},
				pattern color = orange
				]   coordinates {(6,8.90293) (7,18.7653) };
				\addplot [draw = black,
				semithick,
				fill = black
				]   coordinates {(6,5.09497) (7,13.2337)  };
				\addplot [draw = blue,
				semithick,
				fill = blue
				]   coordinates {(6,3.32001) (7,4.71865)  };
			\end{axis}
		\end{tikzpicture}
		\caption{$p^3$ FR with AMR}
		\label{subfig:vp25_3d_comp_p3}
	\end{subfigure}
	\hfill
	\begin{subfigure}[b]{0.45\textwidth}
		\begin{tikzpicture}	
			\begin{axis}[
				xlabel = $D_{max}$,
				ylabel = $\frac{\text{Adaptation}}{\text{Total cost}}$ (\%),
				ytick = {0,1,2,3,4,5,10,15,20},
				xtick = {6,7,8},
				legend style={at={(0,1)},anchor=north west,nodes={scale=0.8, transform shape}}
				]
				\addplot[blue, mark=square*] table {
					6   19.24
					7   10.81
					8   4.19
				};
				\addlegendentry{$p^1$ FR, AMR}
				
				\addplot[red,  mark=triangle*] table {
					6   11.6
					7  4.48
					8  1.94
				};
				\addlegendentry{$p^2$ FR, AMR}
				
				\addplot[black, mark=otimes*] table {
					6   7.63
					7  2.82
				};
				\addlegendentry{$p^3$ FR, AMR}
				
			\end{axis}
		\end{tikzpicture}
		\caption{$D_{max}$ vs. $\frac{\text{Adaptation}}{\text{Total cost}}$ }
		\label{subfig:vp25_3d_speed_ecnt}
	\end{subfigure}
	
	\caption{Computation cost of different components of adaptive $p^1$, $p^2$, and $p^3$ FR on domain $[-12.5,12.5]^2\times[0,1.5625]$.}
\end{figure}
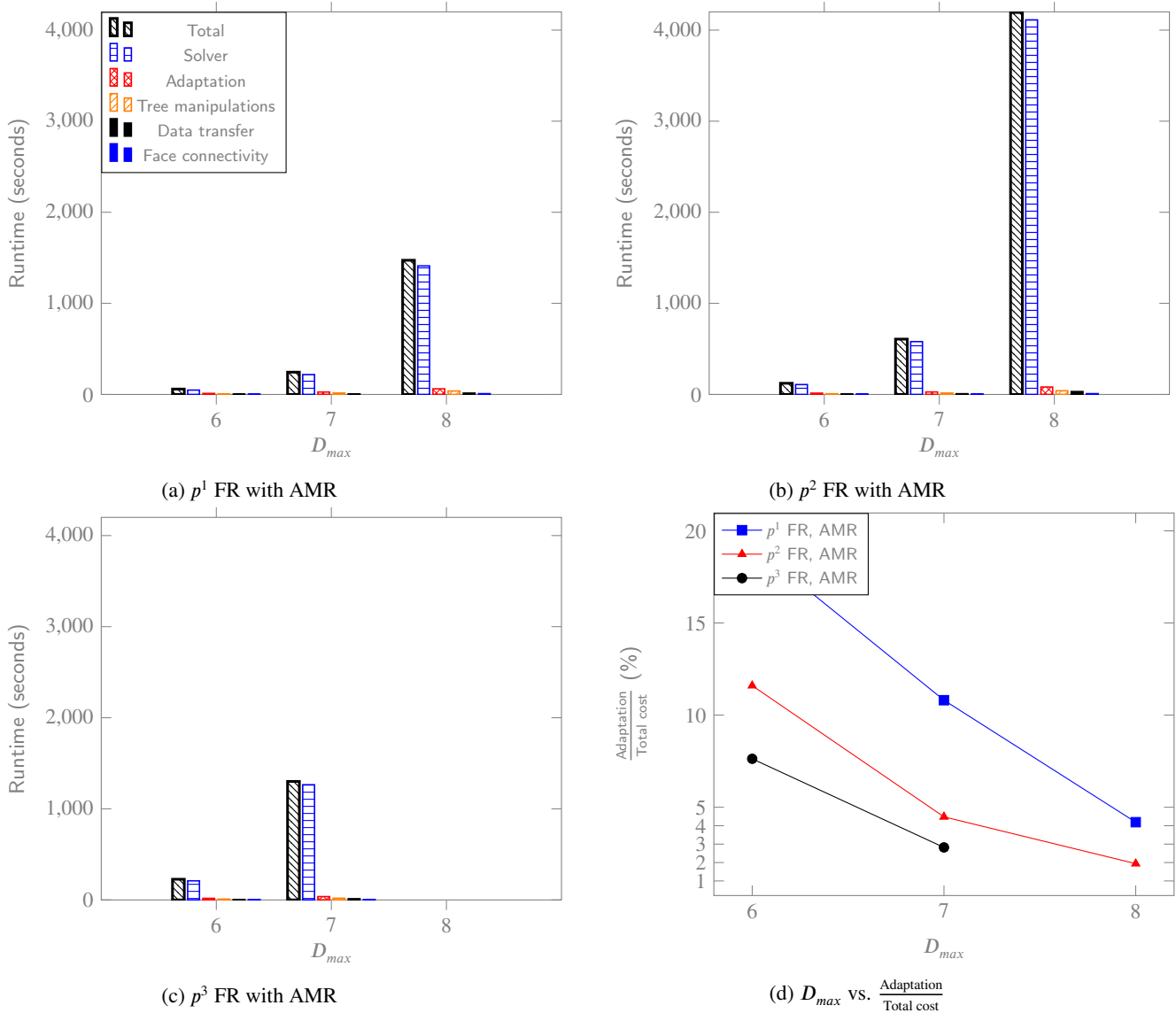

The third tests were conducted on 3D meshes. Simulations were all run with adaptation turned on. The domain on the $xy$ plane was $[-12.5,12,5]^2$ and the thickness of the domain in $z$ direction was $1.5625$.
We showcase a quarter of the vortex and domain after one period with mesh turned on in Figure~\ref{fig:vp_3d_mesh}. 
The 3D effect would lead to a significant element count in the vortex region as we refine the mesh.
The minimum tree depth was set as $D_{min} = 4$ such that in the coarsest region, there was only be one element in $z$ direction.
The maximum tree depth $D_{max}$ varied from 6 to 8 for this test. We used the same adaptation criteria as that in 2D simulations.
In Table~\ref{tab:vp_3d_nonuniform}, we present the overall statistics. The system ran out of memory when the number of element were around $160,000$ for $p^3$ FR. In Figure~\ref{subfig:vp25_3d_comp_p1},~\ref{subfig:vp25_3d_comp_p2}, and~\ref{subfig:vp25_3d_comp_p3}, the computational cost of different components of the entire solver is illustrated. For 3D simulations, the overhead of adaptation compared to the total cost is significantly smaller. Similar to 2D results, tree manipulations took most of the time in adaptation. In Figure~\ref{subfig:vp25_3d_speed_ecnt}, we observe that as the total element count increases or the polynomial degree increases, the ratio $\frac{\text{Adaptation}}{\text{Total cost}}$ will decrease. For $p^2$ FR with $D_{max} = 8$, it is slightly less than 2\%; and for $p^3$ FR with $D_{max} = 7$, it is slightly smaller than 3\%. Compared to low-order methods or 2D problems, more inner points are nested within one element such that the element count becomes relatively smaller compared to the total number of solution points. This further demonstrates that with high-order methods, we can minimize the overhead of adaptation on the GPU hardware. 
\begin{table}[]
	\centering
	\caption{Grid refinement study on domain $[-12.5,12.5]^2\times[0,1.5625]$ with 3D nonuniform meshes.}
	\label{tab:vp_3d_nonuniform}
	\begin{tabular}{lllllll}
		\hline
		\multicolumn{7}{l}{Adaptive $p^1$ FR}\\
		\hline
		$\Delta x_{min}$&  $L_2(\rho)$   & Order  & Total cost (s)  & Adaptation (s) & Element count &$D_{max}$\\
		$25/64$   & $2.19e-02$  &        & 59.05   & 11.36&3049   & 6\\
		$25/128$  & $3.35e-03$  & 2.71   & 244.39  & 26.43&20703  &7\\
		$25/256$  & $5.01e-04$  & 2.74   & 1473.02 & 61.72&160220 &8\\
		\hline
		\multicolumn{7}{l}{Adaptive $p^2$ FR}\\
		\hline
		$\Delta x_{min}$&  $L_2(\rho)$ & Order  & Total cost (s)  & Adaptation (s) & Element count &$D_{max}$ \\
		$25/64$   & $5.37e-04$  &        & 122.56  & 14.22& 3028 &6\\
		$25/128$  & $7.32e-05$  & 2.88   & 607.23  & 27.23& 20759 &7\\
		$25/256$  & $1.07e-05$  & 2.77   & 4193.54 & 81.39& 160346& 8\\
		\hline
		\multicolumn{7}{l}{Adaptive $p^3$ FR}\\
		\hline
		$\Delta x_{min}$&  $L_2(\rho)$ & Order  & Total cost (s)  & Adaptation (s) & Element count & $D_{max}$ \\
		$25/64$   & $1.53e-05$    &        & 226.83  & 17.32& 3028 & 6\\
		$25/128$  & $ 6.52e-07$   & 4.55   & 1301.09 & 36.72&20787 & 7\\
		\hline
	\end{tabular}
\end{table}

\section{Conclusion}\label{sec:summary}
We develop the first entirely GPU-based adaptive flux reconstruction method enabled by linear trees for 2D and 3D problems. Novel algorithms are proposed for tree construction, tree balancing as well as connectivity query. The entire solver runs on GPU except that CPU codes are used to launch the kernel functions. We first show that the GPU version of tree balancing can be 4 to 5 times faster than the parallel CPU version. We demonstrate that significant accelerations for CFD simulations can be achieved for 2D and 3D simulations. An astonishing speedup of 49 can be achieved for long distance vortex propagation compared to uniform meshing. We systematically analyze the computational cost of different components of the entire solver. It is found that the overhead of adaptation, including tree adaptation, tree balancing, mapping, data transferring, face connectivity query, is insignificant compared to the total cost. For 3D simulations, $\frac{\text{Adaptation}}{\text{Total cost}}$ can be as small as  2\%. Overall, high-order methods with efficiently AMR can be achieved entirely operating on GPU. Additionally, high-order methods tend to benefit more from adaptaiton compared to low-order ones. This paves the way for our further development of an automated Cartesian-grid-based CFD solver for multi-GPU platforms. We are to address the initial meshing and wall boundary treatment in our future work as well as extending the developed methods to support multiple GPUs.

\section*{Acknowledgments}
The first author would like to thank Dr. Qimei Gu for her emotional support during the preparation of this manuscript.

\section*{Funding}
The authors declare that no funds, grants, or other support were received during the preparation of this manuscript.
\section*{Competing Interests }
The authors declare that they have no known competing financial interests or personal relationships that could have appeared to influence the work reported in this paper.

\printcredits
\clearpage
\bibliographystyle{unsrtnat}
\bibliography{gpu-based-fr-with-linear-tree}

\end{document}